\documentstyle[12pt,axodraw]{article}
\newcommand{\gx}[2]{\gamma_{#1}\otimes\xi_{#2}}

\newcommand{\ii}[2]{1\otimes1}
\newcommand{\ogx}[2]{\overline{\gamma_{#1}\otimes\xi_{#2}}}

\newcommand{\oii}[2]{\overline{1\otimes1}}
\newcommand{\oogx}[2]{\overline{\overline{\gamma_{#1}\otimes\xi_{#2}}}}

\newcommand{\ooii}[2]{\overline{\overline{1\otimes1}}}
\newcommand{\iiii}[4]{(          1\otimes1       )(          1\otimes1       )}
\newcommand{\oogxgx}[6]{(\overline{\overline{\gamma_{#1}\otimes\xi_{#2}}})_{#5} 
                        (\overline{\overline{\gamma_{#3}\otimes\xi_{#4}}})_{#6}}
\newcommand{\pref}[1]{(\ref{#1})}
\newcommand{\ct}{\bar{c}}
\newcommand{\st}{\bar{s}}
\begin{document}
\title{
\vspace{-3.0cm}
\begin{flushright}
{\normalsize UTHEP-261 }\\
\vspace{-0.3cm}
{\normalsize July 1993 }\\
\end{flushright}
\vspace{2.0cm}
Perturbative renormalization factors for bilinear and four-quark
operators for Kogut-Susskind fermions on the lattice} 
\author{N. Ishizuka\thanks{present address: 
                              Institute of Physics, 
                              University of Tsukuba, 
                              Tsukuba, Ibaraki 305, Japan} \\
        National Laboratory for High Energy Physics(KEK) \\
        Tsukuba, Ibaraki 305, Japan \\
        \and
        Y. Shizawa \\
        Institute of Physics  \\
        University of Tsukuba \\
        Tsukuba, Ibaraki 305, Japan \\
}
\date{}
\maketitle

\begin{abstract}
Renormalization factors for bilinear and four-quark operators with the
Kogut-Susskind fermion action are perturbatively calculated to one-loop order
in the general covariant gauge.
Results are presented both for gauge invariant and non-invariant operators.
For four-quark operators the full renormalization
matrix for a complete set of operators with two types of color contraction
structures are worked out and detailed numerical tables are given.
\end{abstract}

\newpage

%
%
\section{Introduction}

Calculation of weak matrix elements of hadrons is one of the central
subjects in numerical simulation of lattice QCD.
An important ingredient in such studies is the value of renormalization factors
that relate operators on the lattice to those in the continuum.   
In this article we present a one-loop evaluation of renormalization factors 
for bilinear and four-quark operators for the Kogut-Susskind(KS) fermion action.
The results have been used in our recent work for the pion decay constant\cite{fimou}
and the $K^0-\bar K^0$ mixing matrix\cite{ifmosu}.

Perturbative calculation of renormalization factors for KS fermions has been
developed in several previous studies\cite{sharatchandra,goltermansmit,danielsheard,sheard,sharpe}.
In particular Daniel and Sheard\cite{danielsheard,sheard} calculated 
the renormalization factors for KS bilinear operators\cite{danielsheard}
and a subset of four-quark operators\cite{sheard} in the Feynman gauge.
For applications in numerical simulations, however,
their results need to be extended in several directions.
(1)
Weak operators for KS fermions are generally extended in space and time.
The calculation of Daniel and Sheard has been carried out for the operators
which are made gauge invariant through insertion of gauge link variables
between quark and anti-quark fields.
In recent numerical simulations, however,
an alternative method of evaluating matrix elements of gauge non-invariant operators
without link insertions on gauge fixed configurations has been employed\cite{kilcup}.
In fact, whether the two types of operators yield consistent results is the
question we have recently addressed\cite{fimou,ifmosu}.
Analyzing this problem requires the renormalization factors for gauge non-invariant operators 
as well as for gauge invariant ones.
We have therefore carried out the calculation for both types of operators.
(2)
Four-quark operators of form
${\cal O}_2=(\bar q_1^a q_2^a)(\bar q_3^b q_4^b)$ with $a, b$ the color indices generally mix
with those of form
${\cal O}_1=(\bar q_1^a q_2^b)(\bar q_3^b q_4^a)$ under renormalization.
We evaluate the full renormalization factor for the two sets of operators,
while the previous work of Sheard\cite{sheard} listed explicit results only for ${\cal O}_2$
mixing with itself and with ${\cal O}_1$.
(3)
Renormalization factors for lattice operators generally take larger values than
those for continuum operators due to contribution of gluon tadpoles.
Lepage and Mackenzie\cite{lepagemackenzie} have argued
that the tadpole contributions can be removed by a rescaling of quark and gluon fields.
We work out the renormalization factors for the rescaled operators and examine to
what extent their values are reduced by rescaling.
(4)
In addition to the extensions above we have carried out the calculation
in the general covariant gauge which allows us to check the gauge parameter
independence of the results for gauge invariant operators.

For bilinear quark operators a calculation similar to ours has been reported recently
by Patel and Sharpe\cite{patelsharpe}.
Our results are in agreement with theirs and also with those of Ref.~\cite{danielsheard}
for gauge invariant operators.
Patel and Sharpe have extended their calculation to four-quark operators\cite{patelsharpeb}.
For the gauge non-invariant operators which are relevant for the $K$ meson $B$ parameter
their values fully agree with our results.
We also find agreement with the results of Sheard\cite{sheard} for gauge invariant four-quark operators
when a comparison is possible.

We should mention that we do not treat penguin operators in this article. 
Calculation of their renormalization factors 
is technically feasible, which should be pursued in future investigations.

This paper is organized as follows.
In Sec.~2 KS operators whose renormalization factors
we evaluate are defined and a general strategy for one-loop calculations
is summarized following the method of Daniel and Sheard.\cite{danielsheard,sheard}
Results for quark bilinear operators are given in Sec.~3.
Those for four-quark operators are described in Sec.~4 where the relation
between lattice and continuum operators is illustrated for the case of
$\Delta S=2$ operator relevant for extraction of the $K^0-\bar K^0$ mixing matrix.
Analytical expressions for one-loop amplitudes are summarized in Appendices A, B and C.

%
%
\section{Formalism}

\subsection{Quark operators}

The KS fermion action is given by 
\begin{eqnarray}
& &  S =  a^4 \sum_n
    \biggl[  \frac{1}{2a} \sum_\mu 
             \eta_{\mu}(n) \left( \bar{\chi}(n) U_\mu (n) \chi (n + \hat{\mu} ) - 
                             \bar{\chi}(n+\hat{\mu}) U_\mu^{\dag}(n) \chi(n) \right) \cr
& & \qquad\qquad\qquad\qquad\qquad\qquad\qquad\qquad\qquad\qquad
     +  m \bar{\chi}(n) \chi(n) \biggr] \ \ \ ,
\label{dummyishi}
\end{eqnarray}
where $a$ is the lattice spacing, $U_\mu (n)$ denotes the gauge link variable, 
$\eta_\mu (n)=(-1)^{n_1+\cdots +n_{\mu-1}}$ and $\chi$ and $\bar\chi$ 
are the single component KS fermion fields.
For the construction of four-component Dirac fields we employ the coordinate-space method
of taking a linear combination of the $\chi$ fields
over a hypercube\cite{gliozzi,klubergsternetal}.
The Dirac field $Q(2N)$ defined for each hypercube $2N\in (2Z)^4$ is given by
\begin{equation} 
  Q(2N)_{\alpha i} = {1\over 8} \sum_A \left( \gamma_A \right)_{\alpha i} \chi (2N+A)
  \ \ \  , 
\label{eq:quark}
\end{equation} 
where $\alpha$ and $i$ are the Dirac and flavor indices, and 
\begin{equation} 
   \gamma_A=\gamma_1^{A_1} \gamma_2^{A_2} \gamma_3^{A_3}\gamma_4^{A_4} 
\end{equation}
with $A$ running over the vertices of a hypercube ({\it i.e.,} $A_\mu =$0 or 1, $\mu=1,\cdots, 4$).

The bilinear quark operator we use is defined by
\begin{equation}
    {\cal O}_{SF} = \bar{Q} ( \gx{S}{F} ) Q \ \ \ .
\label{eq:bilinearQ}
\end{equation}
Here $\gamma_S=\gamma_1^{S_1} \gamma_2^{S_2} \gamma_3^{S_3} \gamma_4^{S_4}$ and 
$\xi_F=\gamma_1^{*F_1} \gamma_2^{*F_2}\gamma_3^{*F_3} \gamma_4^{*F_4}$ 
with the components $S_\mu$ and $F_\mu$ either 0 or 1.
They act on spinor and flavor indices of the Dirac field $Q$, and represent the spin and flavor
$SU(4)$ content of the bilinear operator.
In terms of the $\chi$ field this operator can be written as  
\begin{equation}
    {\cal O}_{SF} = {1 \over 16} \sum_{AB} \bar{\chi}_{A} ( \ogx{S}{F} )_{AB} \chi_{B} \ \ \ ,
\label{eq:bilinearQchi}
\end{equation}
where we write $\chi_A$ instead of $\chi (2N+A)$ for simplicity and 
\begin{equation}
  (\ogx{S}{F})_{AB} = 
  \frac{1}{4} {\rm tr}(\gamma_A^{\dag} \gamma_S \gamma_B \gamma_F^{\dag}) \ \ \ .
\end{equation}

The operator (\ref{eq:bilinearQchi}) is not gauge invariant.
In order to make it gauge invariant we insert the average of products
of gauge link variables along all possible shortest paths 
connecting the cite $2N+A$ and $2N+B$.
Denoting the link factor by $U_{AB}$ we then define the gauge invariant bilinear
operator by
\begin{equation}
 {\cal O}_{SF}= {1 \over 16}  \sum_{AB}
     \bar{\chi}_{A}^a ( \ogx{S}{F} )_{AB} U_{AB}^{ab} \chi_{B}^b \ \ \ ,
\label{eq:bilinearQinv}
\end{equation}
with $a$ and $b$ the color indices.

We consider two types of four-quark operators differing in the
color contraction structure defined by  
\begin{equation}
 {\cal O}_2 =
  \bar{Q}^{a} ( \gx{S_1}{F_1} ) Q^{a} \cdot
  \bar{Q}^{b} ( \gx{S_2}{F_2} ) Q^{b} \ \ \ , 
\end{equation} 
\begin{equation}
 {\cal O}_1 =
   \bar{Q}^{a} ( \gx{S_1}{F_1} ) Q^{b} \cdot
   \bar{Q}^{b} ( \gx{S_2}{F_2} ) Q^{a} \ \ \ , 
\end{equation} 
with $a$ and $b$ the color indices.
In the method of Ref.~\cite{kilcupetal}
for calculating weak matrix elements with KS fermions the operator ${\cal O}_2$
yields an amplitude with two color contractions after fermion integration and
${\cal O}_1$ with a single color contraction.
For this reason we shall refer to ${\cal O}_2$ and ${\cal O}_1$ as color two-loop
and one-loop operators.
These two operators generally mix under renormalization.

The expression of the operators above in terms of the $\chi$ field is given by 
\begin{eqnarray}
& &
  {\cal O}_2 = \left( {1 \over 16} \right)^2
  \sum_{ABCD} \bar{\chi}^a_A ( \ogx{S_1}{F_1} )_{AB} \chi^a_B \cdot
              \bar{\chi}^b_C ( \ogx{S_2}{F_2} )_{CD} \chi^b_D \ \ \ , \\
& &
  {\cal O}_1 = \left( {1 \over 16} \right)^2
  \sum_{ABCD} \bar{\chi}^a_A ( \ogx{S_1}{F_1} )_{AB} \chi^b_B \cdot
              \bar{\chi}^b_C ( \ogx{S_2}{F_2} )_{CD} \chi^a_D \ \ \ ,
\end{eqnarray}
where we again ignored the hypercube label $2N$ in $\chi(2N+A)$  for simplicity.
To make these operators gauge invariant we insert gauge link factors according to 
\begin{eqnarray}
& &
 {\cal O}_2 = \left( {1\over 16} \right)^2 
   \sum_{ABCD} \bar{\chi}^a_A ( \ogx{S_1}{F_1} )_{AB} \chi^b_B \cdot
               \bar{\chi}^c_C ( \ogx{S_2}{F_2} )_{CD} \chi^d_D \cdot 
        U_{AB}^{ab} U_{CD}^{cd} \ \ \ , \\ 
& &
 {\cal O}_1 = \left( {1\over 16} \right)^2
   \sum_{ABCD} \bar{\chi}^a_A ( \ogx{S_1}{F_1} )_{AB} \chi^b_B \cdot 
               \bar{\chi}^c_C ( \ogx{S_2}{F_2} )_{CD} \chi^d_D \cdot
        U_{AD}^{ad} U_{CB}^{cb} \ \ \ .
\end{eqnarray}
For the color two-loop operators the length of paths for the link factors
$U_{AB}$ and $U_{CD}$ is fixed since $( \ogx{S}{F} )_{AB}$ 
is non-vanishing only for $A+B=S+F({\rm mod} 2)$.
On the other hand, the length varies for the factor
$U_{AD}$ and $U_{CB}$ that appear in the color one-loop operator.

%
%
\subsection{Feynman rules}

We adopt the general covariant gauge with a gauge parameter $\alpha$
in our perturbative calculations.
The gluon propagator is given by  
\begin{equation}
 D_{\mu\nu}(k)_{IJ} = 
   {  
      { \delta_{IJ} \delta_{\mu\nu} }
    \over 
      { \sum_\beta {4 \over a^2} \sin^2 (a k_\beta / 2) }  
   }
 - ( 1- \alpha )
      {
         { \delta_{IJ} \cdot 
           {4 \over a^2 } \cdot \sin (a k_\mu / 2) \cdot \sin (a k_\nu / 2) }
      \over 
         { \left[ \sum_\beta {4 \over a^2} \sin^2 ( a k_\beta / 2 ) \right ]^2 }
      } \ \ \ ,
\end{equation}
and the KS fermion propagator takes the form 
\begin{equation}
 S(p,-q)_{ab} = \delta_{ab} 
  {     \sum_\mu \frac{-i}{a} \sin a p_\mu \bar{\delta}( p - q + \pi\bar{\mu}/a ) +
        m \bar{\delta}( p - q ) 
    \over 
        \sum_\mu \frac{1}{a^2} \sin^2 a p_\mu + m^2
  } \ \ \ ,
\end{equation}
where 
\begin{equation}
 \bar{\mu} = \sum_{\nu=1}^{\mu-1} \hat{\nu} \ \ \ , 
\end{equation}
and
\begin{equation}
 \bar{\delta}(p) = ( 2 \pi )^4 \sum_{n} \delta ( p + \frac{2\pi}{a}n) \ \ \ .
\end{equation}

The one-gluon vertex arising from the action is given by 
\begin{equation}
V_\mu(p,-q;k)_{ab} = 
 -i g (T^I)_{ab} \cdot \cos a (p + k/2 )_\mu 
 \cdot \bar{\delta} ( p - q + k + \pi \bar{\mu} /a ) \ \ \ ,
\end{equation}
with $T^I$ the $SU(3)$ generators normalized by 
${\rm Tr} (T^I T^J)=\delta_{IJ}/2$, and the two-gluon vertex by
\begin{eqnarray}
V_{\mu\nu}(p,-q;k_1,k_2)_{ab}=
 i a g^2 \frac{1}{2} \{T^I,T^J\}_{ab} \cdot \sin a (p+\frac{k_1+k_2}{2})_\mu \cr 
 \times \delta_{\mu\nu} \bar{\delta}( p - q + k_1 + k_2 + \pi \bar{\mu} /a ) \ \ \ ,
\end{eqnarray}
where $p$, $q$ are the incoming fermion momenta and $k$, $k_1$, $k_2$ the gluon momenta.

Vertices of bilinear operators (\ref{eq:bilinearQinv}) have the form,
\begin{eqnarray}
& &M^{(0)}_{SF}(p,-q)_{ab} = \delta_{ab} 
             { 1 \over 16 }  \sum_{AB} e^{iap \cdot A - iaq \cdot B} (\ogx{S}{F})_{AB} \ \ \ , \\
& & M^{(1)}_{SF}(p,-q;k)_{ab} = -iga (T^I)_{ab} \cr
& & \qquad\qquad\qquad
    \times { 1 \over 16 } \sum_{AB} 
     e^{iap\cdot A - iaq \cdot B} (\ogx{S}{F})_{AB} 
                     (A-B)_\mu f^{\mu}_{(AB)}(ak) \ \ \ , \\
& & M^{(2)}_{SF}(p,-q;k_1,k_2)_{ab} = \frac{1}{2}(iga)^2 \frac{1}{2} \{T^I,T^J\}_{ab} \cr 
& & \quad \times { 1 \over 16 } \sum_{AB} e^{iap \cdot A-iaq \cdot B}
    (\ogx{S}{F})_{AB} (A-B)_\mu (A-B)_\nu  g^{\mu\nu}_{(AB)}(ak_1,ak_2) \ \ \ ,
\end{eqnarray}
where the superscript in parentheses denote the number of emitted gluons.
The function $f^\mu_{(AB)}(ak)$ is defined by 
\begin{equation}
 f^\mu_{(AB)}(ak)= e^{iaA\cdot k} {1\over 12} 
   \sum_{\nu \not= \mu} \sum_{j=1}^4 e^{i(B-A) \cdot \theta_{\mu\nu}^{(j)} (ak)}  
\end{equation}
with 
\begin{equation}
\begin{array}{ll}
   \displaystyle
   \theta_{\mu\nu}^{(1)} (ak) = {1 \over 2} ak_\mu \hat{\mu}
   \quad ,
 & \quad 
   \displaystyle
   \theta_{\mu\nu}^{(2)} (ak) = {1 \over 2} ak_\mu \hat{\mu} + ak_\nu \hat{\nu} \\
   \displaystyle
   \theta_{\mu\nu}^{(3)} (ak) = \sum_{\rho=1}^{4} a k_\rho \hat{\rho} - \theta_{\mu\nu}^{(1)}(ak)
   \quad ,
 & \quad 
   \displaystyle
   \theta_{\mu\nu}^{(4)} (ak) = \sum_{\rho=1}^{4} a k_\rho \hat{\rho} - \theta_{\mu\nu}^{(2)}(ak) \\
\end{array} \ \ \ .
\end{equation}
At the one-loop level the two-gluon vertex appears only through gluon tadpole diagrams.
Thus we only need the expression for an equal color index $a=b$
and for the gluon momenta $k_1=-k_2\equiv k$.
In this case we find
\begin{eqnarray}
& & g^{\mu\nu}_{(AB)}(ak,-ak) \cr
& & \quad   = 1 \qquad {\rm for}\, \mu=\nu                     \cr
& & \quad   = e^{iak \cdot ( \Delta_\mu + \Delta_\nu )}
              \left[ 6 + 2 \sum_{\rho\not= \mu\nu} e^{iak\cdot\Delta_\rho }
                       + 2 e^{iak\cdot\sum_{\rho\not= \mu\nu} \Delta_\rho }
              \right] + {\rm h.c } \qquad {\rm for}\,\, \mu\ne\nu     \cr
& &
\end{eqnarray}
with $\Delta_\mu=(B-A)_\mu \hat{\mu}$.

Vertices for four-quark operators are given by
a product of those for the bilinear operators except that the color and site
indices have to be interchanged appropriately.  

%
%
\subsection{Procedure of calculation} 

Let us consider a one-loop diagram of a bilinear operator with two external fermion lines.
The corresponding amplitude generated by the Feynman rules above
are written in terms of momenta $p$ taking values in the range $-\pi/a < p\leq\pi/a$.
Since the Dirac field $Q(2N)$ is defined on sites with even
coordinates the physical momentum $\tilde p$ for quarks is related to $p$
through $p=\tilde p + C \pi/a$ where the vector $C$ ($C_\mu=0$ or 1)
represents the spin-flavor content of quarks.
We extract renormalization factors from Feynman amplitudes evaluated
at vanishing physical momenta $\tilde p=0$ for external fermion lines.
We therefore set the external fermion momenta to $p=C \pi/a$ and $D \pi/a$.
In this case the spin-flavor part of the
tree amplitude $M^{(0)}_{SF}$ takes the form
\begin{equation}
(\oogx{S}{F})_{CD} = {1 \over 16} \sum_{AB} (-1)^{A \cdot C+B \cdot D} (\ogx{S}{F})_{AB} \ \ \ ,
\label{eq:doulebarbasis}
\end{equation}
which shows that the calculation of renormalization factors  requires a
conversion of spin-flavor Dirac structure from the \lq single-bar\rq\ basis,
in which the Feynman rules are given, to the \lq double-bar\rq\ basis defined
in (\ref{eq:doulebarbasis}).  

We employ the technique developed by Daniel and Sheard\cite{danielsheard,sheard}
to carry out the conversion.  
The general form of one-loop amplitudes that results is given by
\begin{equation}
  \sum_{MNM'N'} \int^{\pi/a}_{-\pi/a} { d^4 k a^4 \over (2\pi)^4} \cdot
   A_{MNM'N'}(ak) (\oogx{MSN}{M'FN'})_{CD}
  \ \ \ , 
\label{eq:amplitude}
\end{equation}
where $A_{MNM'N'}$ is a function of loop momenta $k$.
The product of Dirac matrices $(\oogx{MSN}{M'FN'})$ can be reexpanded in terms of the basis
$\{(\oogx{S}{F}); S_\mu, F_\mu = 0, 1\}$;
\begin{equation}
   (\oogx{MSN}{M'FN'}) = \sum_{S'F'} C_{MNM'N'}^{SFS'F'} \cdot (\oogx{S'}{F'}) \ \ \ .
\label{eq:expand}
\end{equation}
The contribution of ({\ref{eq:amplitude}) to the renormalization factor of 
${\cal O}_{SF}$ is  given by  
\begin{equation}
 \sum_{MNM'N'} C_{MNM'N'}^{SFS'F'} \int^{\pi/a}_{-\pi/a}
               \frac{d^4 k a^4 }{(2\pi)^4} A_{MNM'N'}(ak) \ \ \ .
\label{eq:sum}
\end{equation}
In most cases the decomposition (\ref{eq:expand}) is too tedious to work out analytically.
We generate tables of $C_{MNM'N'}^{SFS'F'}$ on a computer and combined them
with tables of one-loop integrals $A_{MNM'N'}$,
separately evaluated with the Monte Carlo integration routine VEGAS,
to calculate the sum (\ref{eq:sum}).  

Our calculations are carried out for massless quark.
In this case the amplitudes for the diagrams which have counterparts
in the continuum perturbation theory contain infrared divergent terms of form
\begin{equation}
  \int_{-\pi/a}^{\pi/a} {d^4 k  a^4 \over (2\pi)^4 }
      { 1 
        \over 
        [ \sum_\mu 4 \sin^2 ( k_\mu a / 2) ] \cdot [ \sum_\nu 4 \sin^2 ( k_\nu a / 2) ]
      }  \ \ \ ,
\end{equation} 
where the first factor in the denominator arises from the gluon propagator and
the second from the massless quark propagator.
To regularize the divergence we supply a finite mass $\kappa$ to the gluon propagator.
The integral then takes the value 
\begin{eqnarray}
 & &  \int_{-\pi/a}^{\pi/a} { d^4 k a^4 \over (2\pi)^4 } 
       { 1 
         \over \left[ \sum_{\mu} 4 \sin^2(k_\mu a / 2)               \right] \cdot 
               \left[ \sum_{\nu} 4 \sin^2(k_\nu a / 2) + (a\kappa)^2 \right]  
       }  \cr
 & & \cr
 & &  = {1 \over 16\pi^2 }
        \left[ -2\cdot \log(a\kappa) + F_{0000} - \gamma_E + 1 \right] + O(\kappa a)
     \ \ \  ,     
\label{IRR:dviterm}
\end{eqnarray}
with $F_{0000}=4.36923(1)$ and $\gamma_E=0.577216\cdots$.

The infrared regularization above is different from that of Daniel and Sheard
who added the mass term $(a \kappa )^2$ to both the quark and gluon propagators,
in which case the finite part of the integral is given by $F_{0000} - \gamma_E$.
We prefer not to adopt their regulator since it leads to a violation of fermion number conservation
in continuum perturbation theory.
We also note that the dependence on the gluon mass should cancel out
between the renormalization factors in the continuum and on
the lattice as long as one employs the same infrared regularization in the two cases.

Evaluation of one-loop amplitudes for four-quark operators are much more
cumbersome since they contain a product of two spin-flavor Dirac matrices. 
The calculational procedure, however, is essentially the same as for bilinear operators.

%
%
\section{Bilinear operators} 

The one-loop renormalization of the quark bilinear operator ${\cal O}_{SF}$
on the lattice can be written as
\begin{equation}
  {\cal O}_{SF}^{lat\  (1)} 
= \sum_{S'F'} \big( \delta_{SS'}\delta_{FF'} 
+  {g^2\over 16\pi^2}  z^{lat}_{SF,S'F'}\big) {\cal O}^{lat\  (0)}_{S'F'}
 \ \ \ ,  
\end{equation}
where the superscript $j$  on ${\cal O}_{SF}^{lat\  (j)}$ refer to the number of loops.
The one-loop diagrams are shown in Fig.~\ref{bilineargp}
and analytic expressions of the amplitudes are collected in Appendix A.
Defining the coefficient $z^{cont}$ for the continuum operators in a similar manner the
one-loop relation between the lattice and continuum operators are given by
\begin{equation}
   {\cal O}^{cont\ (1)}_{SF} = 
 \sum_{S'F'} 
    \left[ \delta_{SS'}\delta_{FF'} + {g^2\over 16\pi^2} ( z^{cont}_{SF,S'F'}-z^{lat}_{SF,S'F'}) 
    \right]
      {\cal O}^{lat\ (1)}_{S'F'} \ \ \ . 
\label{fromlattocont}
\end{equation}

The continuum renormalization factor for massless quarks can be expressed as
\begin{equation}
 z^{cont}_{SF,S'F'} = \delta_{SS'} \delta_{FF'} \cdot \gamma_S \log(\mu/\kappa) 
                    + \delta_{SS'} \delta_{FF'} C_{S}^{cont} \ \ \ .
\label{contref}
\end{equation}
We used the same infrared regularization as for the lattice, 
{\em i.e.} a finite gluon mass $\kappa$ is given to the gluon propagator.
The anomalous dimension of the operator $\gamma_S$ is given by  
\begin{equation}
   \gamma_S = {8 \over 3} \cdot ( \sigma_S -1 ) \ \ \ ,
\end{equation}
with $\sigma_S = (4,1,0,1,4)$ for the spin structure
$\gamma_{S}=(I$, $\gamma_\mu$, $\gamma_{\mu\nu}$, $\gamma_{\mu 5}$, $\gamma_5)$.
The finite constant $C_{S}^{cont}$ depends on the continuum
regularization and renormalization schemes.
For the $\overline{\rm MS}$ subtraction scheme it takes the values 
\begin{equation}
   C_S^{cont} = \left\{
\begin{array}{rrrrrl}
     ( 10/3, & 0, &  2/3, & 0, & 10/3 ) & \qquad \hbox{for NDR}                     \cr
     ( 14/3, & 0, & -2/3, & 0, & 14/3 ) & \qquad \hbox{for DR}\overline{\rm EZ}
\end{array}
   \right.
\label{contr}
\end{equation}
for $\gamma_S=(I$, $\gamma_\mu$, $\gamma_{\mu\nu}$, $\gamma_{\mu 5}$, $\gamma_5)$,
where NDR refers to the naive dimensional regularization with an
anti-commuting $\gamma_5$ and DR$\overline{\rm EZ}$ to the variant of the
dimensional reduction defined in Ref.~\cite{drez}.

For the renormalization factor on the lattice we find
\begin{equation}
z^{lat}_{SF,S'F'}=
  - \delta_{SS'} \delta_{FF'} \cdot  \gamma_S \log(a\kappa) +
    C_{SF,S'F'}^{lat} \ \ \ . \label{latref}
\end{equation}
The logarithmically divergent term arises from the diagrams in 
Fig.~\ref{bilineargp}(a) and (d).
It takes the same form for gauge invariant and non-invariant operators,
and is independent of the gauge parameter $\alpha$.
Comparing \pref{contref} and \pref{latref} 
we find that the gluon mass $\kappa$ cancels out between $z^{cont}_{SF,S'F'}$
and $z^{lat}_{SF,S'F'}$ as it should be.

The finite coefficient $C_{SF,S'F'}^{lat}$ has the following properties.
(1) The coefficients have the same value for the two operators
with the spin-flavor structure $(\gx{S}{F})$ and $(\gx{S5}{F5})$ ( see Ref.~\cite{patelsharpe} ).
(2) Our explicit calculation shows that the Landau gauge part of 
$C_{SF,S'F'}^{lat}$ coming from the $(\sin k^\mu/2 \cdot \sin k^\nu/2)$ term of the
gluon propagator is diagonal in spin and flavor.
Their values are the same for the gauge invariant and non-invariant operators.
(3) The remaining part of the coefficient generally mixes different spin-flavor structures.
Chiral $U(1)$ symmetry of the KS action, however,
places a restriction that operators of even distance with
$\sum_{\mu=1}^{4} (S_\mu + F_\mu)$(mod 2) $= 0, 2, 4$ do not mix with those having odd
distance ($\sum_{\mu=1}^{4} (S_\mu + F_\mu )$(mod 2) $= 1, 3$). 
In fact there are only a few non-vanishing off-diagonal elements
(see Table.~\ref{bilinear}(b) below).
Furthermore their values are the same for gauge invariant and non-invariant operators. 

Numerical values of the diagonal constants $C_{SF,SF}^{lat}$ are tabulated in 
Table.~\ref{bilinear}(a) for both gauge invariant (first column) and
non-invariant (second column) operators.
The results for the gauge non-invariant operators are given in the Landau gauge.
The non-vanishing off-diagonal elements are tabulated in Table.~\ref{bilinear}(b).
Numerical accuracy is within 0.1\%. 
For the gauge invariant operators we have numerically checked the independence of
results on the gauge parameter $\alpha$.
Our results confirm those of Daniel and Sheard ( Table 4 and 5 in Ref.~\cite{danielsheard} )
after correcting our $C_{SF,S'F'}^{lat}$ by $-\delta_{SS'}\delta_{FF'}\gamma_{S}/2$
to take into account the difference in the regularization of infrared divergence.  
They are in a complete agreement with the recent calculation of the same quantity
reported by Patel and Sharpe ( Table 6 and 7 in Ref.~\cite{patelsharpe} ).
The relationship between our coefficients $C_{SF,S'F'}^{lat}$ and theirs $C_{SF,S'F'}^{\rm PS}$ is
\begin{equation}
  C_{SF,S'F'}^{\rm PS} = {3 \over 4} 
            \left[ 
               \delta_{SS'} \delta_{FF'} 
               \left( 
                  C_{S}^{cont} + \gamma_S \log\pi 
               \right)
               - C_{SF,S'F'}^{lat}
            \right] \ \ \ ,
\end{equation}
where $C_{S}^{cont}$ are the finite continuum renormalization factor
for the DR$\overline{\rm EZ}$ scheme given by (\ref{contr}).

We observe in Table.~\ref{bilinear}(a) that the coefficients for
gauge non-invariant operators have a similar magnitude while those for gauge
invariant ones show a substantial variation from operator to operator.
This stems from the fact that the gauge invariant operators receive contribution of
gluon tadpoles whose magnitude increases with the distance of the operator
$\Delta_{SF}=\sum_\mu ( S_\mu + F_\mu )$(mod 2).
Lepage and Mackenzie\cite{lepagemackenzie} have argued that the tadpole contributions
can be removed by a rescaling of fields which, for KS fermion action, takes the form
\begin{equation}
      \chi      \to \sqrt{u_0} \chi       \ \ ,
\quad \bar \chi \to \sqrt{u_0} \bar{\chi} \ \ ,
\quad U_{\mu}   \to u_0^{-1} U_{\mu}      \ \ ,
\label{eq:rescaling}
\end{equation}
where $u_0$ represents the tadpole renormalization of link variables.
A gauge-invariant choice for $u_0$ is given by
\begin{equation}
 u_0 = \left[{1\over 3} \langle {\rm Tr}\ U_P \rangle \right]^{1/4} = 1 - {1 \over 12} g^2 + O(g^4)
\label{eq:tadpole}
\end{equation}
with $\langle {\rm Tr}\ U_P \rangle$ the plaquette average.
For gauge-invariant bilinear operators the rescaling amounts 
to a multiplication by a factor $u_0^{1- \Delta_{SF}}$.
The renormalization factors for the rescaled operators are obtained by subtracting
$4\pi^2(1-\Delta_{SF})/3$ from the second column of Table.~\ref{bilinear}(a).
The results listed in the third column of Table.~\ref{bilinear}(a) show that rescaled
gauge invariant operators receive much less renormalization,
and that their magnitude becomes less dependent on the flavor of operators.
For gauge non-invariant operators without insertion of link variables the rescaling factor
is universally given by $u_0$.
The rescaling reduces the magnitude of the correction
without spoiling the weak flavor dependence already apparent for the original operators. 

%
%
\section{Four-quark operators \label{fourfelmiop} }
\subsection{Lattice result}

One-loop diagrams which contribute to the renormalization of
the color two-loop four-quark operators ${\cal O}_2$ defined in Sec.~2
are shown in Fig.~\ref{fourfermigp}
and those for color one-loop operators ${\cal O}_1$ in Fig.~\ref{fourfermigpone}.
In these figures horizontal lines at the four-quark vertices signify contraction
of spin-flavor quantum numbers, while dotted lines represent link factors and flow of color indices.  
 
For the diagrams of Fig.~\ref{fourfermigp}(a)--(e) for the color two-loop operator evaluation
of momentum and Dirac matrix parts are the same as those of the diagrams in
Fig.~\ref{bilineargp} for the bilinear operator.
The color factor is also the same for these diagrams.
For the diagrams of Fig.~\ref{fourfermigp}(f)--(h), on the other hand,
the color factor takes the form $\sum_I(T^I)_{ab}(T^I)_{a'b'}$
which has to be decomposed into the color one- and two-loop basis.
This can be done by the $SU(3)$ identity, 
\begin{equation} 
  \sum_I (T^I)_{ab} (T^I)_{a'b'} =  -\frac{1}{6}
  \delta_{ab}\delta_{a'b'} + \frac{1}{2} \delta_{ab'}\delta_{a'b} \ \ \ . 
\end{equation}
Such a rearrangement is also generally needed for the diagrams
of color one-loop operators in Fig.~\ref{fourfermigpone}
in addition to manipulation of momentum and Dirac parts.
Analytic expressions for all the diagrams are summarized in Appendices B and C.  

Due to the mixing of color one- and two-loop operators the renormalization
factor for the four-quark operators ${\cal O}_i$ takes a $2\times 2$ matrix form,
\begin{equation}
 {\cal O}^{lat\ (1)}_{i}=\left( \delta_{ij} + {g^2\over 16\pi^2} z^{lat}_{ij} \right)
 {\cal O}^{lat\ (0)}_{j} \ \ , \qquad i,j=1,2 \ \ \ .
\label{fourfermiop:lattice}
\end{equation}
Since the operators
${\cal O}_i = \bar{Q}(\gamma_{S_1}\otimes \xi_{F_1})Q \cdot
              \bar{Q}(\gamma_{S_2} \otimes \xi_{F_2} ) Q$
further depend on the pair of spin-flavor indices $sf\equiv (S_1F_1)(S_2F_2)$,
each element $z_{ij}^{lat}$ is a matrix $z_{ij}^{lat} = \{z_{ij;sf,s'f'}\}$.
Treating the infrared divergence as in the case of bilinear operators
we find that this matrix can be written as
\begin{equation}
 z^{lat}_{ij;sf,s'f'} = - \delta_{ff'} \gamma_{ij;ss'}^{lat} \cdot
    \log(a\kappa) + C_{ij;sf,s'f'}^{lat}
    \ \ \ .
\label{eq:fourquarkz}
\end{equation}

In Table~\ref{table:v5v5:a5a5:inv}--\ref{table:v0a5:a0v5:noninv} we list the numerical values of
the matrices $\gamma_{ij;ss'}^{lat}$ and $C_{ij;sf,s'f'}^{lat}$ for the operators
with the spin-flavor structure
\begin{equation}
\begin{array}{lllll}
 sf = & (\gamma_\mu     \otimes \xi_{5})(\gamma_\mu     \otimes \xi_{5}) 
      & , 
      & (\gamma_{\mu 5} \otimes \xi_{5})(\gamma_{\mu 5} \otimes \xi_{5})
      & \quad  \mbox{(Table 2 and 6)}\ \ ,  \cr
      & (\gamma_{\mu}   \otimes \xi_{5})(\gamma_{\mu 5} \otimes \xi_{5}) 
      & ,
      &                                                                  
      & \quad  \mbox{(Table 3 and 7)}\ \ ,  \cr
      & (\gamma_{\mu}   \otimes       I)(\gamma_\mu     \otimes \xi_{5}) 
      & , 
      &                                                                  
      & \quad  \mbox{(Table 4 and 8)}\ \ ,  \cr
      & (\gamma_{\mu}   \otimes       I)(\gamma_{\mu 5} \otimes \xi_{5}) 
      & ,
      & (\gamma_{\mu 5} \otimes       I)(\gamma_{\mu}   \otimes \xi_{5}) 
      & \quad  \mbox{(Table 5 and 9)}\ \ .
\end{array}
\end{equation}
The results for the gauge invariant operators are
in Table.~\ref{table:v5v5:a5a5:inv}--\ref{table:v0a5:a0v5:inv}
and those for the gauge non-invariant operators are
in Table~\ref{table:v5v5:a5a5:noninv}--\ref{table:v0a5:a0v5:noninv}.
The anomalous dimension matrix $\gamma_{ij;ss'}^{lat}$
is gauge independent and takes the same value for gauge invariant and non-invariant operators.
The results of $C_{ij;sf,s'f'}^{lat}$ for the gauge non-invariant operators are for the Landau gauge.

The one-loop renormalization coefficients for the operators having the spin-flavor structure
$(\gamma_{S5} \otimes \xi_{F5})(\gamma_{S'5}\otimes \xi_{F'5})$
are the same as those for
$(\gamma_S\otimes \xi_F)(\gamma_{S'}\otimes \xi_{F'})$.
Hence the tables also cover the renormalization factor for the operators obtained by the interchange
$\gamma_\mu \leftrightarrow \gamma_{\mu 5}$, $I \leftrightarrow \xi_5$.
These operators are the most relevant for calculation of matrix elements of the effective weak Hamiltonian.
The gauge non-invariant operators with the spin-flavor structure
$(\gamma_{S5} \otimes \xi_{F5})(\gamma_{S'}\otimes \xi_{F'})$
are renormalized in the same way as those for
$(\gamma_S\otimes \xi_F)(\gamma_{S'}\otimes \xi_{F'})$ at least to one loop order, and similarly for the 
operators $(\gamma_S\otimes \xi_F)(\gamma_{S'5}\otimes \xi_{F'5})$ and 
$(\gamma_S\otimes \xi_F)(\gamma_{S'}\otimes \xi_{F'})$\cite{private}.
It is not known if this property persists at higher orders.
We also note that the operators of even distance do not mix with those of odd distance
due to $U(1)$ chiral symmetry, similar to the case of bilinear operators. 

The numerical accuracy is within the level of $\pm 0.001$ for the majority of elements in the tables,
increasing to $\pm 0.01$ for large elements whose magnitude is $O(10)$.
This accuracy should be sufficient for practical applications.
Reducing errors is quite computer time consuming because of a very large number
of lattice integrals ($\sim 360$) which have to be computed.

We have checked the results in two ways.
(i) For gauge invariant operators the Landau gauge part proportional to $1-\alpha$ has to vanish.
This has been confirmed numerically.
(ii) One can rewrite the color one-loop operator as a linear combination of color two-loop operators
through the Fierz transformation given by    
\begin{equation}
 (\ogx{S}{F})_{AB} (\ogx{S'}{F'})_{A'B'} = 
 \frac{1}{16} \sum_{DE} 
    (\overline{\gamma_S \gamma_D^{\dag} \otimes \xi_E^{\dag} \xi_{F'}})_{AB'}
    (\overline{\gamma_{S'} \gamma_D     \otimes \xi_E           \xi_F})_{A'B} \ \ \ .
\label{eq:fierz}
\end{equation}
The results for the color two-loop operators can then be used to evaluate the
renormalization factors for the color one-loop operators.
The results obtained in this way agree with those of a direct calculation
of the color one-loop operators.

The matrix elements for the mixing of gauge invariant color two-loop operators
with color one- and two-loop operators have been computed previously by Sheard\cite{sheard}.
He took a basis quite different from ours,
and we found it difficult to make a full comparison. 
For those cases where we can compare, however, our results are in agreement with his results.  
Also the results in Table.~\ref{table:v5v5:a5a5:noninv} for gauge non-invariant operators agree  
with those of Patel and Sharpe\cite{patelsharpeb}.

Let us finally consider improvement of four-quark operators by factoring
out tadpole renormalizations through rescaling of fields as suggested
by Lepage and Mackenzie\cite{lepagemackenzie}.
For the color two-loop operators the rescaling (\ref{eq:rescaling}) yields a simple form
for the improved operators,
\begin{equation}
   {\cal O}_{2;sf}^{imp}=u_0^{2-\Delta_{sf}}{\cal O}_{2;sf}
\end{equation}
where $\Delta_{sf}=\Delta_{ {S}_1 {F}_1 } + \Delta_{ {S}_2 {F}_2 }$ for gauge invariant operators 
having the spin-flavor structure 
$sf=(\gamma_{S_1}\otimes \xi_{F_1})(\gamma_{S_2}\otimes \xi_{F_2})$ with 
$\Delta_{SF}=\sum_{\mu} (S_\mu + F_\mu)$(mod 2) the distance of bilinear operators, while 
$\Delta_{sf}=0$ for the gauge non-invariant operators.
For the color one-loop operators, on the other hand, rescaling is not straightforward since the link 
insertion factors have lengths ranging from 0 to 4.
To handle this case we recall the Fierz formula (\ref{eq:fierz}) and rewrite the color one-loop
operators in terms of color two-loop operators as
\begin{equation}
  {\cal O}_{1; sf} = \sum_{s'f'} F_{sf,s'f'}\ {\cal O}_{2; s'f'} \ \ \ ,
\end{equation}
with $F_{sf,s'f'}$ numerical constants.
The rescaled color one-loop operators can then be defined as\cite{private}
\begin{equation}
 {\cal O}_{1; sf}^{imp} = \sum_{s'f'} F_{sf,s'f'}\ u_0^{2-\Delta_{s'f'}}\ {\cal O}_{2; s'f'} \ \ \ .
\end{equation} 
For the gauge invariant choice (\ref{eq:tadpole}) for the tadpole factor $u_0$  
the finite renormalization factors for rescaled four-quark operators are listed
in Table.~\ref{table:v5v5:a5a5:inv}--\ref{table:v0a5:a0v5:noninv}
where those elements changed by rescaling are given after a slush $(/)$ symbol.  
As one can see in the tables the rescaling indeed reduces the magnitude of the renormalization correction.

%
%
\subsection{Relation with continuum operators}

In order to obtain the physical values of weak matrix elements the
renormalization factor on the lattice obtained in the previous section has to be
combined with those in the continuum.
In this section we illustrate the procedure
for the $K$ meson $B$ parameter relevant for the $K^0-\bar K^0$ mixing matrix.

The $K$ meson $B$ parameter $B_K$ in the continuum theory is defined by
\begin{equation}
     B_K = {
             \langle \bar{K}^0 | \bar{s} \gamma_\mu (1-\gamma_5) d
                                 \bar{s} \gamma_\mu (1-\gamma_5) d | K^0 \rangle
                \over 
             { 8 \over 3 }f_K^2m_K^2
           } \ \ \ .
\label{bparam}
\end{equation}
In the method of Ref.~\cite{kilcupetal} for calculating weak matrix elements with KS fermions,
the operator in the numerator is replaced by the sum of the following four operators 
\begin{eqnarray}
     & &{\cal V}_1 = \bar{S}^a (\gamma_\mu     \otimes \xi_5) D^b \cdot
                     \bar{S}^b (\gamma_\mu     \otimes \xi_5) D^a\ \ , \cr
     & &{\cal V}_2 = \bar{S}^a (\gamma_\mu     \otimes \xi_5) D^a \cdot
                     \bar{S}^b (\gamma_\mu     \otimes \xi_5) D^b\ \ , \cr
     & &{\cal A}_1 = \bar{S}^a (\gamma_{\mu 5} \otimes \xi_5) D^b \cdot
                     \bar{S}^b (\gamma_{\mu 5} \otimes \xi_5) D^a\ \ , \cr
     & &{\cal A}_2 = \bar{S}^a (\gamma_{\mu 5} \otimes \xi_5) D^a \cdot 
                     \bar{S}^b (\gamma_{\mu 5} \otimes \xi_5) D^b\ \ ,
\label{fourop}
\end{eqnarray}
where $S$ and $D$ are the KS quark fields introduced for $s$ and $d$ quark
separately, $a$ and $b$ the color indices, and the quark fields in the first
current are to be contracted with $\bar{K}^0$ and those in the second current with $K^0$.
The choice of flavor $\xi_5$ in these operators corresponds
to the use of $\bar{D} ( \gamma_5 \otimes \xi_5) S$ for creating the external $K^0$ and $\bar K^0$
in the Nambu-Goldstone channel associated with $U(1)$ chiral symmetry of the KS fermion action.

The renormalization factor for these operators can be read off from
Table.~2 for gauge invariant operators and from Table.~6 for gauge non-invariant operators.
As can be seen, the four operators not only mix among themselves but
also with a large number of others having different spin-flavor structures.
We note that the extra operators all have the flavor matrix $\xi_F\ne\xi_5$.
Since the $K^0$ and $\bar K^0$ mesons are created with the flavor $\xi_5$,
the matrix element of the extra operators should vanish in the continuum limit where
a restoration of $SU(4)$ flavor symmetry is expected.
The renormalization factor for some of the extra operators are numerically not small, however.
Whether they yield negligibly small contributions at the current
range of inverse lattice spacing $1/a\sim 2-3$GeV has to be checked through actual simulations.
For simplicity we disregard the mixing with the extra operators in the following.
Extensions to the general case is straightforward.

Denoting the four operators \pref{fourop} as 
$\{{\cal O}_\alpha^{lat};\alpha=1,\cdots,4\}=\{{\cal V}_1,\cdots,{\cal A}_2\}$, 
we find that the $4\times 4$ anomalous dimension matrix
$\gamma_{\alpha\beta}^{lat}$ is given by
\begin{equation}
\gamma_{\alpha\beta}^{lat}=\left(
      \begin{array}{rrrr}
              9  &  -3  &  -7  & -3 \\ 
              0  &   0  &  -6  &  2 \\ 
             -7  &  -3  &   9  & -3 \\ 
             -6  &   2  &   0  &  0 \\ 
           \end{array} 
     \right) \ \ \ ,
\end{equation}
which takes the same form for the gauge invariant and non-invariant cases.  
The finite part $C_{\alpha\beta}^{lat}$ takes the values,  
\begin{equation}
 C_{\alpha\beta}^{lat} = 
   \left(
     \begin{array}{rrrr}
            -18.915 &  -4.772 &  -5.253 & -2.251 \\
                  0 & -60.000 &  -4.502 &  1.501 \\
             -5.253 &  -2.251 & -19.513 & -2.977 \\
             -4.502 &   1.501 &       0 &      0 \\
     \end{array}
   \right)
\label{eq:inv}
\end{equation}
for the gauge invariant operator, and 
\begin{equation}
 C_{\alpha\beta}^{lat} = 
   \left(
     \begin{array}{rrrr}
           37.446 & -2.913 & -5.253 & -2.251 \\
                0 & 28.706 & -4.502 &  1.501 \\
           -5.253 & -2.251 & 37.976 & -4.504 \\
           -4.502 &  1.501 &      0 & 24.464 \\
     \end{array}
   \right)
\label{eq:noninv}
\end{equation}
for the gauge non-invariant operator in the Landau gauge.
The result for the gauge non-invariant operator (\ref{eq:noninv})
is in agreement with those of Ref.~\cite{patelsharpeb}.
The second and the fourth row of (\ref{eq:inv}) has previously been calculated by Sheard,
with which our results agree.
For the rescaled operators discussed in Sec.~4.1
the diagonal elements of $C_{\alpha\beta}^{lat}$ changes to 
( $ 7.403$, $-7.361$, $ 6.805$, $     0$ ) for the gauge invariant operators and to 
( $11.127$, $ 2.387$, $11.657$, $-1.854$ ) for the gauge non-invariant operators
(off-diagonal elements are not affected).

The $\Delta S=2$ continuum operators are given by 
\begin{eqnarray}
& & {\cal L}_1 = \bar{s}^a \gamma_\mu (1-\gamma_5) d^b \cdot
                 \bar{s}^b \gamma_\mu (1-\gamma_5) d^a \ \ \ , \cr
& & {\cal L}_2 = \bar{s}^a \gamma_\mu (1-\gamma_5) d^a \cdot
                 \bar{s}^b \gamma_\mu (1-\gamma_5) d^b \ \ \ ,
\end{eqnarray}
with the one-loop renormalization taking the form
\begin{equation}
  {\cal L}_i^{(1)}
  = \left[ \delta_{ij} + {g^2 \over 16\pi^2}
           \big(\gamma_{ij}^{cont} \cdot \log (\mu /\kappa)  + C_{ij}^{cont}\big)
    \right] {\cal L}_j^{(0)} \ \ \ ,
\label{llopmix}
\end{equation}
where the anomalous dimension matrix $\gamma_{ij}^{cont}$ is given by
\begin{equation}
\gamma_{ij}^{cont} = 
\left(
  \begin{array}{rr}
       2  & -6   \\
      -6  &  2  
  \end{array}
\right)
\ \ \ .
\end{equation}
For the finite part $C_{ij}^{cont}$ we find that
\begin{equation}
   C_{ij}^{cont} = c\gamma_{ij}^{cont}
\end{equation}
with 
\begin{equation}
c= \left\{
\begin{array}{rl}
     11/12 & \qquad \hbox{for NDR} \cite{ndr}    \cr
      7/12 & \qquad \hbox{for DR}\overline{\rm EZ}
\end{array}
   \right. \ \ \ .
\end{equation}

In order to relate the lattice operators to those in the continuum  let us
define a $2\times 4$ matrix $M_{i\alpha}$ by
\begin{equation}
  M = \pmatrix{ 1 & 0 & 1 & 0 \cr 
                0 & 1 & 0 & 1 }
\end{equation}
with which one can write ${\cal L}_i^{(0)}=M_{i\alpha}{\cal O}_\alpha^{lat\ (0)}$.
Using the intertwining property
$M_{i\alpha}\gamma^{lat}_{\alpha\beta}=\gamma^{cont}_{ij}M_{j\beta}$
it is easy to see that the one-loop renormalization relation between the continuum
and lattice operators is given by
\begin{equation}
 {\cal L}_i^{(1)} = M_{i\alpha} 
     \left[ 
          \delta_{\alpha\beta}+\frac{g^2}{16\pi^2}
          \big( \gamma^{lat}\log (\mu a)+c\gamma^{lat} - C^{lat} \big)_{\alpha\beta}
     \right]
 {\cal O}_\beta^{lat\ (1)} \ \ \ .
\end{equation}
The numerator of the $B_K$ parameter with one-loop renormalization
correction equals the sum 
$\langle \bar{K}^0 |{\cal L}_1^{(1)} | K^0 \rangle + \langle \bar{K}^0 |{\cal L}_2^{(1)} | K^0 \rangle$.

%
%
%
\section*{Acknowledgments}
Valuable discussions and correspondence with Steve Sharpe and Aproova Patel are gratefully acknowledged.  
We thank A. Ukawa and M. Okawa for useful discussions,
and M. Fukugita and H. Mino for comments on the manuscript.

%
\newpage
\appendix
\section*{Appendix A One-loop amplitudes for bilinear operators}

The one-loop amplitudes corresponding to the diagrams of Fig.~\ref{bilineargp}
for the external quark momenta $p=C \pi/a$, $D \pi/a$
and color indices $a$, $b$ are as follows(the common factor
$4/3\cdot \delta_{ab} \cdot g^2/(16\pi^2)$ is not included).
\begin{eqnarray}
     G^{\ref{bilineargp}(a)} 
 &=& \left( \sigma_S  - (1-\alpha) \right) x (\oogx{S}{F})_{CD}
   + \sum_{\mu\rho\sigma MN} X_{MN}^{\mu,\rho\sigma} (\oogx{\mu\rho MSN\sigma\mu}{MFN})_{CD} \cr
 &-& (1-\alpha) \sum_M X_M (\oogx{MSM}{MFM})_{CD} \\
     G^{\ref{bilineargp}(b)}
 &=& \sum_{MN\mu\rho} Y_{MN}^{\mu,\rho} |S+F|_\mu
     (\oogx{\mu\rho\mu 5MSN}{\mu 5MFN} +  \oogx{\mu 5MSN\rho\mu}{\mu 5MFN})_{CD} \cr
 &-& 4(1-\alpha) \sum_{MN\mu} Y_{MN}^\mu |S+F|_\mu
     (\oogx{\mu 5MSN}{\mu 5MFN})_{CD}        \\ 
     G^{\ref{bilineargp}(c)}
 &=& {1\over 8} (3 + \alpha) Z_{0000} (\oogx{S}{F})_{CD}  \\
     G^{\ref{bilineargp}(d)}
 &=&  \left[ - x \alpha  - {1\over 8}(1+\alpha) Z_{0000} - R \right] (\oogx{S}{F})_{CD} \\ 
     G^{\ref{bilineargp}(e)}
 &=& \left( - {1\over 8}(3+\alpha) \Delta_{SF} Z_{0000} - 2(1-\alpha) T_{ \Delta_{SF} }
     \right) (\oogx{S}{F})_{CD}
\end{eqnarray}
where $\sigma_S = (4,1,0,1,4)$ for 
$S=(I,\gamma_\mu,\gamma_{\mu\nu},\gamma_{\mu 5},\gamma_5)$,
$x=-2\log a\kappa + F_{0000} - \gamma_E + 1$,
$|S+F|_\mu = (S_\mu + F_\mu)$ (mod 2), and
$\Delta_{SF}=\sum_\mu |S+F|_{\mu}$.

We use the following notations to simplify the expressions for the loop
integrals which appear in the amplitudes above:
\begin{eqnarray}
& & s_\mu = \sin \phi_\mu\ \ , \qquad \st_\mu = \sin {\phi_\mu /2}\ \ ,
                               \qquad B = [ 4 \sum_\mu \st^2_\mu ]^{-1} \ \ , \cr 
     & & c_\mu = \cos \phi_\mu\ \ , \qquad \ct_\mu = \cos {\phi_\mu /2}\ \ ,
                               \qquad F = [ \sum_\mu s^2_\mu ]^{-1} \ \ , \cr
     & & \int_\phi = 16\pi^2 \int^{\pi}_{-\pi} {d^4\phi \over (2\pi)^4} \ \ ,
\end{eqnarray} 
where $\phi$ is the loop momentum and $B$ and $F$ originate from gluon and fermion propagators.
In terms of these symbols the loop integrals are defined as follows.
\begin{eqnarray}
& &
 X^{\mu,\rho\sigma}_{MN} = \int_\phi
    \biggl[ 
          \ct_\mu^2 s_\rho s_\sigma E_M(\phi) E_N(-\phi) B F^2
       - {1\over 4}\delta_{\rho\sigma} \delta_{M0} \delta_{N0} B^2 
    \biggr]  \\
& & X_M = \int_\phi
    \biggl[ 
            E_M(\phi)E_M(-\phi) - \delta_{M0}
    \biggr] B^2   \\
& &
Y^{\mu,\rho}_{MN} = \int_\phi i \ct_\mu S_\rho B F
   \cdot   {1 \over 12} \sum_{j=1}^4 \sum_{\sigma \not= \mu} 
            E_M( \theta_{\mu\sigma}^{(j)} ) E_N( -\theta_{\mu\sigma}^{(j)} )   \\
& &
Y^{\mu}_{MN} = \int_\phi i \st_\mu B^2
    {1 \over 12} \sum_{j=1}^4 \sum_{\sigma \not= \mu}
            E_M( \theta_{\mu\sigma}^{(j)} ) E_N( -\theta_{\mu\sigma}^{(j)} )   \\
& &
Z_{0000} = \int_\phi B \\
& &
T_{\Delta} = \int_\phi
           2  \st^2_1 \st^2_2 B^2 \times ( 0, 0, 1, 2 + c_3, 3 + 2c_3 + c_3 c_4 ) \cr
& & \qquad\qquad\qquad\qquad\qquad
            \qquad \hbox{for}\quad \Delta = (0, 1, 2, 3, 4)            \\
& &
R  =  \int_\phi
      \left[
          c_1 ( 2 s^2_1 - {1\over F} )( -2 - 2 \st^2_1 + \frac{1}{4 B} )  B F^2 - B^2
      \right]  
\end{eqnarray}
where $E_M(\phi)$ and $\theta_{\mu\nu}$ are given by
\begin{equation}
 E_M( \phi ) = \prod_\mu \frac{1}{2} 
   \left( e^{-i\phi_\mu/2} + (-1)^{\tilde{M}_\mu} e^{i\phi_\mu/2} \right) \ \ \ ,
\end{equation}
\begin{equation} 
  \tilde{M}_\mu=\sum_{\nu\neq\mu} M_\nu \ \ \ ,
\end{equation}
\begin{equation}
\begin{array}{lll}
   \displaystyle
   \theta^{(1)}_{\mu\nu} = {1\over 2} \phi_\mu \hat\nu 
 & , &
    \displaystyle
   \theta^{(2)}_{\mu\nu} = {1\over 2} \phi_\mu \hat\mu + \phi_\nu\hat\nu \\
   \displaystyle
   \theta^{(3)}_{\mu\nu} = \sum_{\rho=1}^4 \phi_\rho \hat{\rho} - \theta^{(1)}_{\mu\nu} 
 & , &
   \displaystyle
   \theta^{(4)}_{\mu\nu} = \sum_{\rho=1}^4 \phi_\rho \hat{\rho} - \theta^{(2)}_{\mu\nu} \\
 & & \\
\end{array} \ \ \ .
\end{equation}

The integrals $X_{MN}^{\mu,\rho\sigma}$ and $Y_{MN}^{\mu\rho}$
were evaluated numerically by Daniel and Sheard\cite{danielsheard}.
We have confirmed their results except for some sign reversals for $Y_{MN}^{11}$ in their Table.~3.
For the calculation of the integral we employed the Monte Carlo integration routine VEGAS.

%
%
\section*{Appendix B One-loop amplitudes for color two-loop four-quark operators}

Analytic expressions for one-loop diagrams in Fig.~\ref{fourfermigp}
are listed below for color two-loop four-quark operators of a general spin-flavor structure
$(\gamma_S \otimes \xi_F)(\gamma_{S'} \otimes \xi_{F'})$.
The amplitudes for the diagrams Fig.~\ref{fourfermigp}(a)--(e) are products of
tree and one-loop bilinear amplitudes.
The diagrams in Fig.~\ref{fourfermigp}(f)--(h) cannot be factorized in this way.
They take the following form, where we drop the common factor
$\sum_I (T^I)_{ab}(T^I)_{a'b'}$ and $g^2/16\pi^2$.
External fermion lines have momenta $p=C \pi/a\cdot$, $D\pi/a$, $C' \pi/a$, $D' \pi/a$
and color indices $a,b,a',b'$ as specified in Fig.~\ref{fourfermigp}.
\newpage
\begin{eqnarray}
\lefteqn{G^{\ref{fourfermigp}(f)} = }  \cr
& & - {1\over 4} x  \sum_{\mu\rho} ( \oogx{\mu\rho S }{F } - \oogx{S \rho\mu}{F } )_{C D }
                                   ( \oogx{\mu\rho S'}{F'} - \oogx{S'\rho\mu}{F'} )_{C'D'}         \cr
& & - \sum_{\mu\rho\sigma MN} X^{\mu,\rho\sigma}_{MN}
                              ( \oogx{\mu\rho   MS }{M F } - \oogx{S  M\rho  \mu}{F M} )_{C D } \cr
& & \qquad\qquad\qquad \times ( \oogx{\mu\sigma MS'}{M F'} - \oogx{S' M\sigma\mu}{F'M} )_{C'D'} \cr
& & + (1-\alpha) \sum_M X_M
   ( \oogx{M S }{M F } - \oogx{S M }{F  M} )_{C D } 
   ( \oogx{M S'}{M F'} - \oogx{S'M }{F' M} )_{C'D'}  \cr
& & \\
\lefteqn{ G^{\ref{fourfermigp}(g)} = } \cr
& & \sum_{\mu\rho LMN} U^{\mu,\rho}_{LMN} \biggl\{ \cr
& & \qquad\quad |S +F |_\mu ( \oogx{\mu 5MSL}{\mu 5MFL})_{CD} 
                  ( \oogx{S'N \rho\mu}{F'N} - \oogx{\mu\rho NS'}{NF'})_{C' D'}  \cr 
& & \qquad +    |S'+F'|_\mu ( \oogx{SN \rho\mu}{FN} - \oogx{\mu\rho NS}{NF})_{CD} 
                  (\oogx{\mu 5MS' L}{\mu 5MF' L})_{C' D'} 
    \biggr\} \cr
& & - (1-\alpha) \sum_{\mu LMN} U^{\mu}_{LMN} \biggl\{ \cr
& &  \qquad\quad 
       |S +F |_\mu ( \oogx{\mu 5MSL}{\mu 5MFL})_{CD} 
                   ( \oogx{S'N}{F'N} - \oogx{NS'}{NF'})_{C' D'} \cr
& &  \qquad
     + |S'+F'|_\mu ( \oogx{SN}{FN} - \oogx{NS}{NF})_{CD} 
                   (\oogx{\mu 5MS' L}{\mu 5MF' L})_{C' D'}  
     \biggr\} \cr
& & \\
\lefteqn{ G^{\ref{fourfermigp}(h)} =  } \cr
& & - \sum_{\mu KLMN} V^\mu_{KLMN} |S+F|_\mu |S'+F'|_\mu 
       (\oogx{\mu 5KS L}{\mu 5KF L})_{C D } (\oogx{\mu 5MS'N}{\mu 5MF'N})_{C'D'} \cr
& & + (1 - \alpha ) \sum_{\mu\nu KLMN} V^{\mu\nu}_{KLMN} 
                                    |S + F|_\mu (\oogx{\mu 5KS L}{\mu 5KF L})_{C D } \cr
& & \qquad\qquad\qquad\qquad\qquad
                             \times |S'+F'|_\nu (\oogx{\nu 5MS'N}{\nu 5MF'N})_{C'D'} \cr
& &
\end{eqnarray}
where $|S + F|_\mu=(S_\mu + F_\mu)$(mod 2).  
The four types of loop integrals are defined by 
\begin{eqnarray}
& & U^{\mu,\rho}_{LMN} = \int_\phi
   i \ct_\mu s_\rho B F 
     {1\over 12} \sum_{\nu\not=\mu} \sum_{j=1}^4 
                   E_L(        \theta_{\mu\nu}^{(j)}) 
                   E_M( \phi - \theta_{\mu\nu}^{(j)})
                   E_N(      - \phi                   )   \\
& & U^{\mu}_{LMN} = \int_\phi
   i 2 \st_\mu B^2
     {1\over 12} \sum_{\nu\not=\mu} \sum_{j=1}^4
                   E_L(        \theta_{\mu\nu}^{(j)})
                   E_M( \phi - \theta_{\mu\nu}^{(j)})
                   E_N(      - \phi                   )   \\
& & V^\mu_{KLMN} =  \int_\phi B
        {1 \over 12} \sum_{ \nu     \neq \mu }  \sum_{ i =1 }^4
            E_K(\theta_{\mu\nu}^{(i)}) E_L( \phi - \theta_{\mu\nu}^{(i)})  \cr
& & \qquad\qquad\qquad
 \times {1 \over 12} \sum_{ \lambda \neq \mu }  \sum_{ j =1 }^4
            E_M(-\theta_{\mu\lambda}^{(j)}) E_N(-\phi + \theta_{\mu\lambda}^{(j)}) \\
& & V^{\mu\nu}_{KLMN} =  \int_\phi 4 \st_\mu \st_\nu B^2
 {1 \over 12} \sum_{\lambda \neq \mu} \sum_{i=1}^{4}
      E_K( \theta_{\mu\lambda}^{(i)}) E_L(  \phi - \theta_{\mu\lambda}^{(i)} ) \cr
& & \qquad\qquad\qquad
 \times
 {1 \over 12} \sum_{\sigma \neq \nu}  \sum_{j=1}^{4}
      E_M(-\theta_{\nu\sigma }^{(j)}) E_N( -\phi + \theta_{\nu \sigma}^{(j)} ) \ \ \ .
\end{eqnarray}
The numerical values of the last three integrals reported in Tables 4
and 5 in Ref.~\cite{sheard} have some minus signs missing.
We have corrected the sign and evaluated the values of $U_{LMN}^{\mu,\rho}$
which are not listed in Ref.~\cite{sheard}.

%
%
\section*{Appendix C One-loop amplitudes for color one-loop four-quark operators}

Analytic expressions for one-loop diagrams in Fig.~\ref{fourfermigpone}
are listed below for color one-loop four-quark operators 
of a general spin structure $(\gamma_S \otimes \xi_F)(\gamma_{S'} \otimes \xi_{F'})$.
We set the momenta of the external fermion lines to
$p=C\pi/a$, $D \pi/a$, $C' \pi/a$, $D' \pi/a$
and assign the color indices $a,b,a',b'$ as shown in Fig.~\ref{fourfermigpone}.
The amplitudes of the diagrams Fig.~\ref{fourfermigpone}(a), (c), and (d)
are the same as those  for the color two-loop operators except for the color factor which takes the form 
 $\sum_I (T^I)_{ab'} (T^I)_{a'b}$ for the diagram (a)
and $\delta_{ab'}\delta_{a'b}$ for the diagrams (c) and (d).
Other amplitudes take the following form, where we drop the common factor $g^2/16\pi^2$.
%
%
\begin{eqnarray}
\lefteqn{ G^{\ref{fourfermigpone}(b)} = { 4 \over 3 } \delta_{ab'}\delta_{a'b} 
    \biggl[    } \cr
& & - \sum_{\mu\rho MN} \frac{1}{2} ( Y^{\mu,\rho}_{M[\mu5N]} + Y^{\mu,\rho}_{N[\mu5M]} ) 
    \bigg\{ \cr
& & \qquad\quad 
               \oogxgx{\mu\rho MS}{MF}{S'N}{F'N}{CD}{C'D'} + \oogxgx{SM \rho\mu}{FM}{NS'}{NF'}{CD}{C'D'} \cr
& & \qquad   + \oogxgx{SN}{FN}{\mu\rho MS'}{MF'}{CD}{C'D'} + \oogxgx{NS}{NF}{S'M \rho\mu}{F'M}{CD}{C'D'} 
    \biggr\} \cr
& & - 4 (1-\alpha) \sum_{\mu M} Y^\mu_{[\mu5M]M} 
        \left( (-1)^{\tilde{M}\cdot (S+F)}
              +(-1)^{\tilde{M}\cdot (S'+F')} 
        \right) \cr
& & \qquad\qquad 
       \times \oogxgx{SM}{FM}{S'M}{F'M}{CD}{C'D'} \biggr] \cr
& & \\
\lefteqn{G^{\ref{fourfermigpone}(e)} = { 4 \over 3 }\delta_{ab'}\delta_{a'b} 
    \biggl[                   } \cr
& & - 2 Z_{0000} \oogxgx{S}{F}{S'}{F'}{CD}{C'D'}  \cr
& & + \sum_\mu \frac{1}{4} Z_{0000} ( (-1)^{(S+F)_\mu}+(-1)^{(S'+F')_\mu} ) 
                    \oogxgx{\mu5S}{\mu5F}{\mu5S'}{\mu5F'}{CD}{C'D'} 
      \cr
& & + (1-\alpha) \frac{1}{2} Z_{0000} \oogxgx{S}{F}{S'}{F'}{CD}{C'D'} \cr
& & - (1-\alpha) \sum_\mu Z_{0000} \frac{1}{2} ( (-1)^{(S+F)_\mu} + (-1)^{(S'+F')_\mu} )
            \oogxgx{\mu 5S}{\mu 5F}{\mu 5S'}{\mu 5F'}{CD}{C'D'} \cr
& & + (1-\alpha) \frac{1}{2} \sum_{\mu\neq\nu, M} T^{\mu\nu}_M
        \biggl\{ \cr
& & \qquad   - ( (-1)^{\tilde{M}\cdot(S+F)} + (-1)^{\tilde{M}\cdot(S'+F')} ) 
                     \oogxgx{MS}{MF}{MS'}{MF'}{CD}{C'D'} \cr
& & \qquad   
         + 2 ( (-1)^{ (S+F)_\mu + \tilde{M}\cdot(S+F)} +(-1)^{(S'+F')_\mu  + \tilde{M}\cdot(S'+F')} ) \cr
& & \qquad\qquad\qquad\qquad\qquad
            \times \oogxgx{\mu5MS}{\mu5MF}{\mu5MS'}{\mu5MF'}{CD}{C'D'} \cr
& & \qquad   - (   (-1)^{( S + F )_\mu +( S + F )_\nu + \tilde{M}\cdot( S + F )} 
                 + (-1)^{( S'+ F')_\mu +( S'+ F')_\nu + \tilde{M}\cdot( S'+ F')} ) \cr
& & \qquad\qquad\qquad\qquad\qquad
            \times \oogxgx{\mu\nu MS}{\mu\nu MF}{\mu\nu MS'}{\mu\nu MF'}{CD}{C'D'}
        \biggr\}
    \bigg] \cr
& & \\
\lefteqn{G^{\ref{fourfermigpone}(f)} = } \cr
\lefteqn{\sum_{\mu\rho\sigma MN}
    ( X^{\mu,\rho\sigma}_{MN} + \frac{x}{4} \delta_{\rho\sigma}\delta_{M0}\delta_{N0} )
    \biggl[ } \cr
& & \quad \frac{4}{3} \delta_{ab'} \delta_{a'b} 
    \biggl\{ \cr 
& & \qquad      \oogxgx{\mu\rho MS}{MF}{S'N \sigma\mu}{F'N}{CD}{C'D'}
              + \oogxgx{SM \rho\mu}{FM}{\mu\sigma NS'}{NF'}{CD}{C'D'} 
    \biggr\} \cr
& & \quad - \sum_I (T^I)_{ab'} (T^I)_{a'b}
    \biggl\{ \cr
& & \qquad   \oogxgx{\mu\rho MS}{MF}{\mu\sigma NS'}{NF'}{CD}{C'D'}
           + \oogxgx{SM \rho\mu}{FM}{S'N \sigma\mu}{F'N}{CD}{C'D'} 
    \biggr\} 
    \bigg] \cr
\lefteqn{ + (1-\alpha) \sum_M ( X_M + x \delta_{M0} ) \oogxgx{SM}{FM}{S'M}{F'M}{CD}{C'D'}
      \biggl[ } \cr
& & \quad - \frac{4}{3} \delta_{ab'} \delta_{a'b}
              ( (-1)^{\tilde{M}\cdot (S+F)} + (-1)^{\tilde{M}\cdot (S'+F')} )
   + \sum_I (T^I)_{ab'} (T^I)_{a'b} ( 1 + (-1)^{\tilde{M}\cdot (S+F+S'+F')} ) 
    \biggr] \cr
& & \\
\lefteqn{G^{\ref{fourfermigpone}(g)}  = \sum_I (T^I)_{ab'} (T^I)_{a'b} 
    \biggl[                  } \cr
& & \sum_{\mu\rho LMN} \frac{1}{2} (U^{\mu,\rho}_{[\mu5L]MN} - U^{\mu,\rho}_{[\mu5M]LN})
    \biggl\{ \cr
& & \qquad  \oogxgx{LSN\rho\mu}{LFN}{S'M}{F'M}{CD}{C'D'}
          + \oogxgx{SM}{FM}{LS'N\rho\mu}{LF'N}{CD}{C'D'} \cr
& & \quad + \oogxgx{\mu\rho NSL}{NFL}{MS'}{MF'}{CD}{C'D'}
          + \oogxgx{MS}{MF}{\mu\rho NS'L}{NF'L}{CD}{C'D'}
    \biggr\} \cr
& & - (1-\alpha) \sum_{\mu LMN} \frac{1}{2} \left(U^\mu_{[\mu5L]MN} - U^\mu_{[\mu5M]LN} \right)
    \biggl\{ \cr
& & \qquad  \oogxgx{SM}{FM}{LS'N}{LF'N}{CD}{C'D'}
          + \oogxgx{MS}{MF}{NS'L}{NF'L}{CD}{C'D'} \cr
& & \quad + \oogxgx{LSN}{LFN}{S'M}{F'M}{CD}{C'D'} 
          + \oogxgx{NSL}{NFL}{MS'}{MF'}{CD}{C'D'}
    \biggr\}
    \biggr] \cr
& & \\
\lefteqn{G^{\ref{fourfermigpone}(h)} = \sum_I (T^I)_{ab'} (T^I)_{a'b} 
    \biggl[                  } \cr
& & \sum_{\mu KLMN} \frac{1}{4} \oogxgx{KSN}{KFN}{MS'L}{MF'L}{CD}{C'D'}
    \biggl\{ \cr
& & \qquad   ( (-1)^{(S+F)_\mu} + (-1)^{(S'+F')_\mu}) V^\mu_{KLMN}
           - ( 1 + (-1)^{(S+F+S'+F')_\mu}) V^\mu_{[\mu5K]L[\mu5M]N} 
    \biggr\} \cr
& & + (1-\alpha) \sum_{\mu\nu KLMN} \frac{1}{4} \oogxgx{LSM}{LFM}{NS'K}{NF'K}{CD}{C'D'}
    \biggl\{ \cr
& & \qquad    V^{\mu\nu}_{K[\mu5L]M[\nu5N]} + V^{\mu\nu}_{L[\mu5K]N[\nu5M]}
            - V^{\mu\nu}_{K[\mu5L]N[\nu5M]} - V^{\mu\nu}_{L[\mu5K]M[\nu5N]}
    \bigg\}
    \biggr] \ \ \ . \cr
& &
\end{eqnarray}
The new integral $T^{\mu\nu}_M$ is defined by
\begin{equation}
  T^{\mu\nu}_M = \int_{\phi} \st_\mu^2 \st_\nu^2 B^2
  \left[  { 1\over 6} \sum_{j=1}^{3} 
          E_M( \psi^{(j)}_{(\mu\nu)} ) E_M( -\psi^{(j)}_{(\mu\nu)} ) 
       + {1 \over 2} \delta_{M0}
  \right]
  \ \ \ ,
\end{equation}
where
\begin{eqnarray}
 & &  \psi^{(1)}_{(\mu\nu)} = \phi_\rho \hat{\rho} \ , \cr
 & &  \psi^{(2)}_{(\mu\nu)} = \phi_\sigma \hat{\sigma} \ , \cr
 & &  \psi^{(3)}_{(\mu\nu)} = \phi_\rho \hat{\rho} + \phi_\sigma \hat{\sigma} \ \ \ ,
\end{eqnarray}
the components $\mu$, $\nu$, $\rho$, and $\sigma$ are all different.

%
%
%
\newpage
\newcommand{\NPB}[3]{{Nucl. Phys.} {\bf B#1} (#2) #3.}
%

%
%
%
\newpage
%
%
%
\textheight=  20cm
\textwidth =  16.5cm
\hoffset   = -1.5cm
\newcommand{\I}{I}
\newcommand{\m}{\mu}
\newcommand{\n}{\nu}
\newcommand{\r}{\lambda}
\newcommand{\k}{\sigma}
\newcommand{\f}{5}
\def\g#1{ \ifx #1\I \I \else \gamma_{#1} \fi }
\def\x#1{ \ifx #1\I \I \else \xi_{#1} \fi }
\newcommand{\op}[4]{(\g{#1}\otimes\x{#2})(\g{#3}\otimes\x{#4})}
\newcommand{\opt}[2]{(\g{#1}\otimes\x{#2})}
\newcommand{\mcc}{ \makebox[1mm]{ } }
\newcommand{\mba}{ \makebox[11mm]{ } }
\newcommand{\mbac}{ \makebox[11mm]{ } }
\newcommand{\mbg}{ \makebox[5mm][r]{$\gamma$} }
\newcommand{\mbf}{ \makebox[11mm][r]{finite} }
\newcommand{\mbgc}{ \makebox[5mm]{ } }
\newcommand{\mbfc}{ \makebox[11mm]{ } }
%
%
%
%
\begin{table}[p]
\caption{
One-loop perturbative corrections for bilinear operators. 
Operators of form $(\gamma_S\otimes \xi_F )$ and $(\gamma_{S5}\otimes \xi_{F5})$
receive the same one-loop corrections.
The components $\mu$, $\nu$, $\rho$ and $\sigma$ are all different and not summed.
Values of anomalous dimension $\gamma_S$ are also listed.
(a) Diagonal elements of the renormalization factor for gauge invariant (first column)
and non-invariant (second column) operators.
The third and fourth column shows the values for rescaled operators as discussed in text.
(b) Off-diagonal elements of the correction which take the same values for gauge
invariant and non-invariant operators.
The first and second columns of operators specifies the row and column of the mixing matrix.
\label{bilinear}  }
\begin{center}
\begin{tabular}{rlrrrrr}
%
(a) \\
    & operator       & $\gamma$& invariant   & non-invariant & inv(rescaled) & non-inv(rescaled)  \cr
\hline
  1 & $\opt{\I}{\I}$     & $ 8$    &  $ 55.585$  & $ 55.585$ & $ 42.426$ & $ 42.426$            \cr
  2 & $\opt{\I}{\f}$     & $ 8$    &  $-47.783$  & $ 12.813$ & $ -8.304$ & $ -0.346$            \cr 
  3 & $\opt{\I}{\m}$     & $ 8$    &  $ 14.844$  & $ 27.077$ & $ 14.844$ & $ 13.918$            \cr
  4 & $\opt{\I}{\m\f}$   & $ 8$    &  $-29.948$  & $ 14.405$ & $ -3.629$ & $  1.246$            \cr
  5 & $\opt{\I}{\m\n}$   & $ 8$    &  $-10.569$  & $ 17.583$ & $  2.589$ & $  4.423$            \cr
  6 & $\opt{\m}{\I}$     & $ 0$    &  $  0.000$  & $ 12.232$ & $  0.000$ & $ -0.927$            \cr
  7 & $\opt{\m}{\f}$     & $ 0$    &  $-30.000$  & $ 14.353$ & $ -3.682$ & $  1.193$            \cr
  8 & $\opt{\m}{\m}$     & $ 0$    &  $ 19.693$  & $ 19.693$ & $  6.533$ & $  6.533$            \cr
  9 & $\opt{\m}{\n}$     & $ 0$    &  $-13.388$  & $ 14.764$ & $ -0.228$ & $  1.605$            \cr
 10 & $\opt{\m}{\n\f}$   & $ 0$    &  $-13.409$  & $ 14.743$ & $ -0.249$ & $  1.584$            \cr
 11 & $\opt{\m}{\m\f}$   & $ 0$    &  $-45.988$  & $ 14.608$ & $ -0.651$ & $  1.448$            \cr
 12 & $\opt{\m}{\m\n}$   & $ 0$    &  $  4.519$  & $ 16.752$ & $  4.519$ & $  3.593$            \cr
 13 & $\opt{\m}{\n\r}$   & $ 0$    &  $-29.651$  & $ 14.702$ & $ -3.332$ & $  1.543$            \cr
 14 & $\opt{\m\n}{\I}$   & $-8/3$  &  $-14.623$  & $ 13.529$ & $ -1.464$ & $  0.369$            \cr
 15 & $\opt{\m\n}{\m}$   & $-8/3$  &  $ -0.428$  & $ 11.804$ & $ -0.428$ & $ -1.355$            \cr
 16 & $\opt{\m\n}{\r}$   & $-8/3$  &  $-29.668$  & $ 14.685$ & $ -3.349$ & $  1.525$            \cr
 17 & $\opt{\m\n}{\m\n}$ & $-8/3$  &  $  7.728$  & $  7.728$ & $ -5.430$ & $ -5.430$            \cr
 18 & $\opt{\m\n}{\m\r}$ & $-8/3$  &  $-14.200$  & $ 13.952$ & $ -1.041$ & $  0.792$            \cr
 19 & $\opt{\m\n}{\r\k}$ & $-8/3$  &  $-45.390$  & $ 15.206$ & $ -5.911$ & $  2.046$            \cr
\hline
\end{tabular}
\end{center}
\vskip 0.5cm
\begin{center}
\begin{tabular}{rllrl}
%
(b) \\
    & operator          & mixed operator    &  \cr
\hline
  9 & $\opt{\m}{\n}$    & $\opt{\m}{\m}$                   & $-4.504$ & \cr
 11 & $\opt{\m}{\m\f}$  & $\opt{\m}{\n\f}$                 & $ 0.860$ & \cr
 13 & $\opt{\m}{\n\r}$  & $\opt{\m}{\m\n},\opt{\m}{\r\m}$  & $ 1.980$ & \cr
 16 & $\opt{\m\n}{\r}$  & $\opt{\m\n}{\m},\opt{\m\n}{\n}$  & $ 0.902$ & \cr
\hline
\end{tabular}
\end{center}
\end{table}
%
%
%
\textheight=  9.0in
\textwidth =  6.5in
\hoffset   =-0.75in
\voffset   =-0.77in
%
\newcommand{\0}{I}
\newcommand{\1}{\mu}
\newcommand{\2}{\nu}
\newcommand{\3}{\rho}
\newcommand{\4}{\sigma}
\newcommand{\5}{5}
\def\g#1{\ifx#1\0 I \else \gamma_{#1} \fi}
\def\x#1{\ifx#1\0 I \else    \xi_{#1} \fi}
\newcommand{\ops}[4]{(\g{#1}\otimes\x{#2})(\g{#3}\otimes\x{#4})}
{\scriptsize
%
\begin{table}[p]
\caption{
Anomalous dimensions and finite corrections for gauge invariant four-quark
operators with the spin-flavor structure
$VV=(\gamma_\mu     \otimes \xi_5)(\gamma_\mu     \otimes \xi_5)$ and
$AA=(\gamma_{\mu 5} \otimes \xi_5)(\gamma_{\mu 5} \otimes \xi_5)$.
Subscripts 1 or 2 attached to the operators refer to the the number of color loops.
Operators listed at the top row mixes with those on the first column with the numerical coefficients
given (a common factor $g^2/16\pi^2$ is removed).
The values after `$/$' are the finite corrections for the rescaled operators. 
The two rows of numerical values for each operator in the first column are
for the color one-loop operator (first row) and for the color two-loop operator (second column).
All indices of operators are summed with different indices not taking equal values.
}
\label{table:v5v5:a5a5:inv}   
\begin{center}
\begin{tabular}{rl @{}r@{ }r@{}l @{}r@{ }r@{}l @{}r@{ }r@{}l @{}r@{ }r@{}l }
  & 
  &          & $VV_1$ &
  &          & $VV_2$ &
  &          & $AA_1$ &
  &          & $AA_2$ & \\
  &  mixed operator             
  & \mbgc\mbg & \mbf & \mba 
  & \mcc\mbg  & \mbf & \mba
  & \mcc\mbg  & \mbf & \mba
  & \mcc\mbg  & \mbf & \mba   \\
\hline 
 1&$\ops{\0}  {\1}  {\0}  {\1}  $&$  $&$   4.725$&        &$  $&$        $&         &$  $&$ -19.114$&$/-5.954$&$  $&$  -5.573$& \\
  &$                            $&$  $&$  -1.575$&        &$  $&$        $&         &$  $&$  -2.412$&         &$  $&$   1.857$& \\
 2&$\ops{\0}  {\1}  {\0}  {\2}  $&$  $&$  -0.624$&        &$  $&$        $&         &$  $&$  -0.485$&         &$  $&$   1.112$& \\
  &$                            $&$  $&$   0.208$&        &$  $&$        $&         &$  $&$  -0.208$&         &$  $&$  -0.370$& \\
 3&$\ops{\0}  {\4\5}{\0}  {\3\5}$&$  $&$  -0.052$&        &$  $&$  -0.235$&         &$  $&$  -0.067$&         &$  $&$        $& \\
  &$                            $&$  $&$  -0.022$&        &$  $&$   0.078$&         &$  $&$   0.022$&         &$  $&$        $& \\
\end{tabular}
\end{center}								        
\end{table}								     
\newpage								        
\begin{table}[p]							        
\begin{center}
\begin{tabular}{rl @{}r@{ }r@{}l @{}r@{ }r@{}l @{}r@{ }r@{}l @{}r@{ }r@{}l }
  &
  & \mbgc\mbgc & \mbfc & \mba 
  & \mcc\mbgc  & \mbfc & \mba 
  & \mcc\mbgc  & \mbfc & \mba 
  & \mcc\mbgc  & \mbfc & \mba \\
26&$\ops{\1\2}{\3}  {\1\2}{\3}  $&$  $&$ -19.434$&$/-6.274$&$  $&$  -5.960$&      &$  $&$   4.145$&       &$  $&$        $& \\
  &$                            $&$  $&$  -3.000$&         &$  $&$   1.986$&      &$  $&$  -1.381$&       &$  $&$        $& \\
27&$\ops{\1\2}{\3}  {\1\2}{\4}  $&$  $&$   0.485$&         &$  $&$   1.944$&      &$  $&$   0.624$&       &$  $&$        $& \\
  &$                            $&$  $&$   0.208$&         &$  $&$  -0.648$&      &$  $&$  -0.208$&       &$  $&$        $& \\
28&$\ops{\1\2}{\2\5}{\1\2}{\1\5}$&$  $&$  -0.003$&         &$  $&$        $&      &$  $&$  -0.052$&       &$  $&$  -0.044$& \\
  &$                            $&$  $&$  -0.168$&         &$  $&$        $&      &$  $&$  -0.022$&       &$  $&$   0.014$& \\
29&$\ops{\1\2}{\2\5}{\1\2}{\2\5}$&$  $&$   0.918$&         &$  $&$        $&      &$  $&$  -0.900$&       &$  $&$  -0.771$& \\
\end{tabular}
\end{center}
\end{table}
\newpage
\begin{table}[p]
\caption{
Renormalization factors for the
gauge invariant operators $VA=(\gamma_\mu \otimes \xi_5)(\gamma_{\mu 5} \otimes \xi_5)$.
Matrix elements are arranged in the same way as in
Table.~{\protect \ref{table:v5v5:a5a5:inv}}.
}
\label{table:v5a5:inv}
\begin{center}
\begin{tabular}{rl @{}r@{ }r@{}l @{}r@{ }r@{}l }
  & 
  &          & $VA_1$ &
  &          & $VA_2$ & \\
  & mixed operator 
  & \mbgc\mbg  & \mbf & \mba   
  & \mcc\mbg   & \mbf & \mba \\
\hline
 1&$\ops{\1}  {\0}  {\1\5}{\0}  $&$  $&$   0.428$&        &$  $&$        $&           \\
  &$                            $&$  $&$  -0.142$&        &$  $&$        $&           \\
 2&$\ops{\1}  {\5}  {\1\5}{\5}  $&$ 9$&$ -19.319$&$/6.998$&$  $&$        $&           \\
  &$                            $&$-3$&$  -3.560$&        &$  $&$ -29.999$&$/ -3.680$ \\
 3&$\ops{\1}  {\1\2}{\1\5}{\1\2}$&$  $&$  -1.289$&        &$  $&$  -1.104$&           \\
  &$                            $&$  $&$  -0.552$&        &$  $&$   0.368$&           \\
\end{tabular}
\end{center}
\end{table}
\newpage
\begin{table}[p]
\caption{
Renormalization factors for the gauge invariant operator
$VV=(\gamma_\mu     \otimes I)(\gamma_\mu     \otimes \xi_5)$.
Matrix elements are arranged in the same way as in
Table.~{\protect \ref{table:v5v5:a5a5:inv}}.
}
\label{table:v0v5:a0a5:inv}
\begin{center}
\begin{tabular}{rl @{}r@{ }r@{}l @{}r@{ }r@{}l @{}r@{ }r@{}l }
  & 
  &          & $VV_1$ & 
  &          & $VV_2$ & \\
  & mixed operator
  & \mbgc\mbg & \mbf & \mba 
  & \mcc\mbg  & \mbf & \mba \\
\hline 
 1&$\ops{\1}  {\0}  {\1}    {\5}    $&$ 9$&$ -19.319$&$/ 6.999$&$  $&$        $&           \\
  &$                                $&$-3$&$  -3.560$&         &$  $&$ -29.999$&$/ -3.680$ \\
 2&$\ops{\1}  {\5}  {\1}    {\0}    $&$  $&$   0.428$&         &$  $&$        $&           \\
  &$                                $&$  $&$  -0.142$&         &$  $&$        $&           \\
 3&$\ops{\1}  {\1\2}{\1}    {\2\4\5}$&$  $&$   0.051$&         &$  $&$        $&           \\
  &$                                $&$  $&$  -0.153$&         &$  $&$        $&           \\
\end{tabular}
\end{center}
\end{table}
\newpage
\begin{table}[p]
\caption{
Renormalization factors for the gauge
invariant operator
$VA=(\gamma_\mu     \otimes I)(\gamma_{\mu 5} \otimes \xi_5)$ and
$AV=(\gamma_{\mu 5} \otimes I)(\gamma_\mu     \otimes \xi_5)$.
Matrix elements are arranged in the same way  as in
Table.~{\protect \ref{table:v5v5:a5a5:inv}}.
}
\label{table:v0a5:a0v5:inv}
\begin{center}
\begin{tabular}{rl @{}r@{ }r@{}l @{}r@{ }r@{}l @{}r@{ }r@{}l @{}r@{ }r@{}l }
  & 
  &          & $VA_1$ & 
  &          & $VA_2$ & 
  &          & $AV_1$ & 
  &          & $AV_2$ &  \\
  & mixed operator
  & \mbgc\mbg & \mbf & \mba
  & \mcc\mbg  & \mbf & \mba
  & \mcc\mbg  & \mbf & \mba
  & \mcc\mbg  & \mbf & \mba  \\
\hline
 1&$\ops{\0}  {\1}  {\5}    {\1\5}  $&$  $&$   1.511$&         &$  $&$        $&   &$  $&$  -0.439$&        &$  $&$  -0.376$&         \\
  &$                                $&$  $&$  -0.872$&         &$  $&$        $&   &$  $&$  -0.188$&        &$  $&$   0.125$&         \\
 2&$\ops{\0}  {\1}  {\5}    {\4\5}  $&$  $&$   0.130$&         &$  $&$        $&   &$  $&$   0.052$&        &$  $&$   0.044$&         \\
  &$                                $&$  $&$  -0.213$&         &$  $&$        $&   &$  $&$   0.022$&        &$  $&$  -0.014$&         \\
 3&$\ops{\0}  {\4\5}{\5}    {\1}    $&$  $&$   0.485$&         &$  $&$   0.416$&   &$  $&$   0.319$&        &$  $&$        $&         \\
  &$                                $&$  $&$   0.208$&         &$  $&$  -0.138$&   &$  $&$   0.706$&        &$  $&$        $&         \\
\end{tabular}
\end{center}
\end{table}
\begin{table}[p]
\begin{center}
\begin{tabular}{rl @{}r@{ }r@{}l @{}r@{ }r@{}l @{}r@{ }r@{}l @{}r@{ }r@{}l } 
  &
  & \mbgc\mbgc & \mbfc & \mba
  & \mcc\mbgc  & \mbfc & \mba
  & \mcc\mbgc  & \mbfc & \mba
  & \mcc\mbgc  & \mbfc & \mba \\   
26&$\ops{\1\2}{\1}  {\1\2\5}{\2\5}  $&$  $&$   0.052$&        &$  $&$  -0.528$&    &$  $&$   0.067$&         &$  $&$        $&       \\
  &$                                $&$  $&$   0.022$&        &$  $&$   0.176$&    &$  $&$  -0.022$&         &$  $&$        $&       \\
27&$\ops{\1\2}{\1}  {\1\2\5}{\4\5}  $&$  $&$        $&        &$  $&$        $&    &$  $&$   0.031$&         &$  $&$        $&       \\
  &$                                $&$  $&$        $&        &$  $&$        $&    &$  $&$  -0.095$&         &$  $&$        $&       \\
28&$\ops{\1\2}{\3}  {\1\2\5}{\2\5}  $&$  $&$        $&        &$  $&$        $&    &$  $&$   0.031$&         &$  $&$        $&       \\
  &$                                $&$  $&$        $&        &$  $&$        $&    &$  $&$  -0.095$&         &$  $&$        $&       \\
29&$\ops{\1\2}{\3}  {\1\2\5}{\3\5}  $&$  $&$  -0.900$&        &$  $&$  -0.771$&    &$  $&$   0.918$&         &$  $&$        $&       \\
\end{tabular}
\end{center}
\end{table}
\newpage
\begin{table}[p]
\caption{
Renormalization factors for the gauge non-invariant operators
$VV=(\gamma_\mu     \otimes \xi_5)(\gamma_\mu     \otimes \xi_5)$ and
$AA=(\gamma_{\mu 5} \otimes \xi_5)(\gamma_{\mu 5} \otimes \xi_5)$ in the Landau gauge.
Matrix elements are arranged in the same way as in
Table.~{\protect \ref{table:v5v5:a5a5:inv}}.
}
\label{table:v5v5:a5a5:noninv}
\begin{center}
\begin{tabular}{rl @{}r@{ }r@{}l @{}r@{ }r@{}l @{}r@{ }r@{}l @{}r@{ }r@{}l } 
  & 
  &          & $VV_1$ &
  &          & $VV_2$ &
  &          & $AA_1$ &
  &          & $AA_2$ & \\
  & mixed operator
  & \mbgc\mbg & \mbf & \mba 
  & \mcc\mbg  & \mbf & \mba
  & \mcc\mbg  & \mbf & \mba
  & \mcc\mbg  & \mbf & \mba  \\
\hline
 1&$\ops{\0}  {\1}  {\0}  {\1}   $&$  $&$   4.725$&         &$  $&$        $&        &$  $&$  -3.774$&         &$  $&$  -3.235$&         \\
  &$                             $&$  $&$  -1.575$&         &$  $&$        $&        &$  $&$  -1.617$&         &$  $&$   1.078$&         \\
 2&$\ops{\0}  {\1}  {\0}  {\2}   $&$  $&$  -0.624$&         &$  $&$        $&        &$  $&$  -0.485$&         &$  $&$  -0.416$&         \\
  &$                             $&$  $&$   0.208$&         &$  $&$        $&        &$  $&$  -0.208$&         &$  $&$   0.138$&         \\
 3&$\ops{\0}  {\4\5}{\0}  {\3\5} $&$  $&$  -0.052$&         &$  $&$  -0.044$&        &$  $&$  -0.067$&         &$  $&$        $&         \\
  &$                             $&$  $&$  -0.022$&         &$  $&$   0.014$&        &$  $&$   0.022$&         &$  $&$        $&         \\
\end{tabular}
\end{center}
\end{table}
\newpage															    
\begin{table}[p]
\caption{
Renormalization factors for the gauge non-invariant operators
$VA=(\gamma_\mu \otimes \xi_5)(\gamma_{\mu 5} \otimes \xi_5)$ in the Landau gauge.
Matrix elements are arranged in the same way as in
Table.~{\protect \ref{table:v5v5:a5a5:inv}}.
}
\label{table:v5a5:noninv}
\begin{center}
\begin{tabular}{rl @{}r@{ }r@{}l @{}r@{ }r@{}l }
  & 
  &          & $VA_1$ &
  &          & $VA_2$ & \\
  & mixed operator
  & \mbgc\mbg & \mbf & \mba 
  & \mcc\mbg  & \mbf & \mba  \\
\hline
 1&$\ops{\1}  {\0}  {\1\5}{\0}  $&$  $&$   0.428$&         &$  $&$        $&         \\
  &$                            $&$  $&$  -0.142$&         &$  $&$        $&         \\
 2&$\ops{\1}  {\5}  {\1\5}{\5}  $&$ 9$&$  37.711$&$/11.392$&$  $&$        $&         \\
  &$                            $&$-3$&$  -3.708$&         &$  $&$  26.585$&$/ 0.266$\\
 3&$\ops{\1}  {\1\2}{\1\5}{\1\2}$&$  $&$  -1.289$&         &$  $&$  -1.104$&         \\
  &$                            $&$  $&$  -0.552$&         &$  $&$   0.368$&         \\
\end{tabular}
\end{center}
\end{table}
\newpage
\begin{table}[p]
\caption{
Renormalization factors for the gauge non-invariant operators
$VV=(\gamma_\mu \otimes I)(\gamma_\mu \otimes \xi_5)$ in the Landau gauge.
Matrix elements are arranged in the same way as in
Table.~{\protect \ref{table:v5v5:a5a5:inv}}.
}
\label{table:v0v5:a0a5:noninv}
\begin{center}
\begin{tabular}{rl @{}r@{ }r@{}l @{}r@{ }r@{}l }
  & 
  &          & $VV_1$ &
  &          & $VV_2$ & \\
  & mixed operator
  & \mbgc\mbg & \mbf & \mba 
  & \mcc\mbg  & \mbf & \mba  \\
\hline
 1&$\ops{\1}  {\0}  {\1}    {\5}    $&$ 9$&$  37.711$&$/11.392$&$  $&$        $&         \\
  &$                                $&$-3$&$  -3.708$&         &$  $&$  26.585$&$/ 0.266$\\
 2&$\ops{\1}  {\5}  {\1}    {\0}    $&$  $&$   0.428$&         &$  $&$        $&         \\
  &$                                $&$  $&$  -0.142$&         &$  $&$        $&         \\
 3&$\ops{\1}  {\1\2}{\1}    {\1\4\5}$&$  $&$  -0.174$&         &$  $&$  -0.149$&         \\
  &$                                $&$  $&$  -0.074$&         &$  $&$   0.049$&         \\
\end{tabular}
\end{center}
\end{table}
\newpage
\begin{table}[p]
\caption{
Renormalization factors  for the gauge non-invariant operators
$VA=(\gamma_\mu     \otimes I)(\gamma_{\mu 5} \otimes \xi_5)$ and
$AV=(\gamma_{\mu 5} \otimes I)(\gamma_\mu     \otimes \xi_5)$ in the Landau gauge.
Matrix elements are arranged in the same way as in
Table.~{\protect \ref{table:v5v5:a5a5:inv}}.
}
\label{table:v0a5:a0v5:noninv}
\begin{center}
%
\end{center}
\caption{
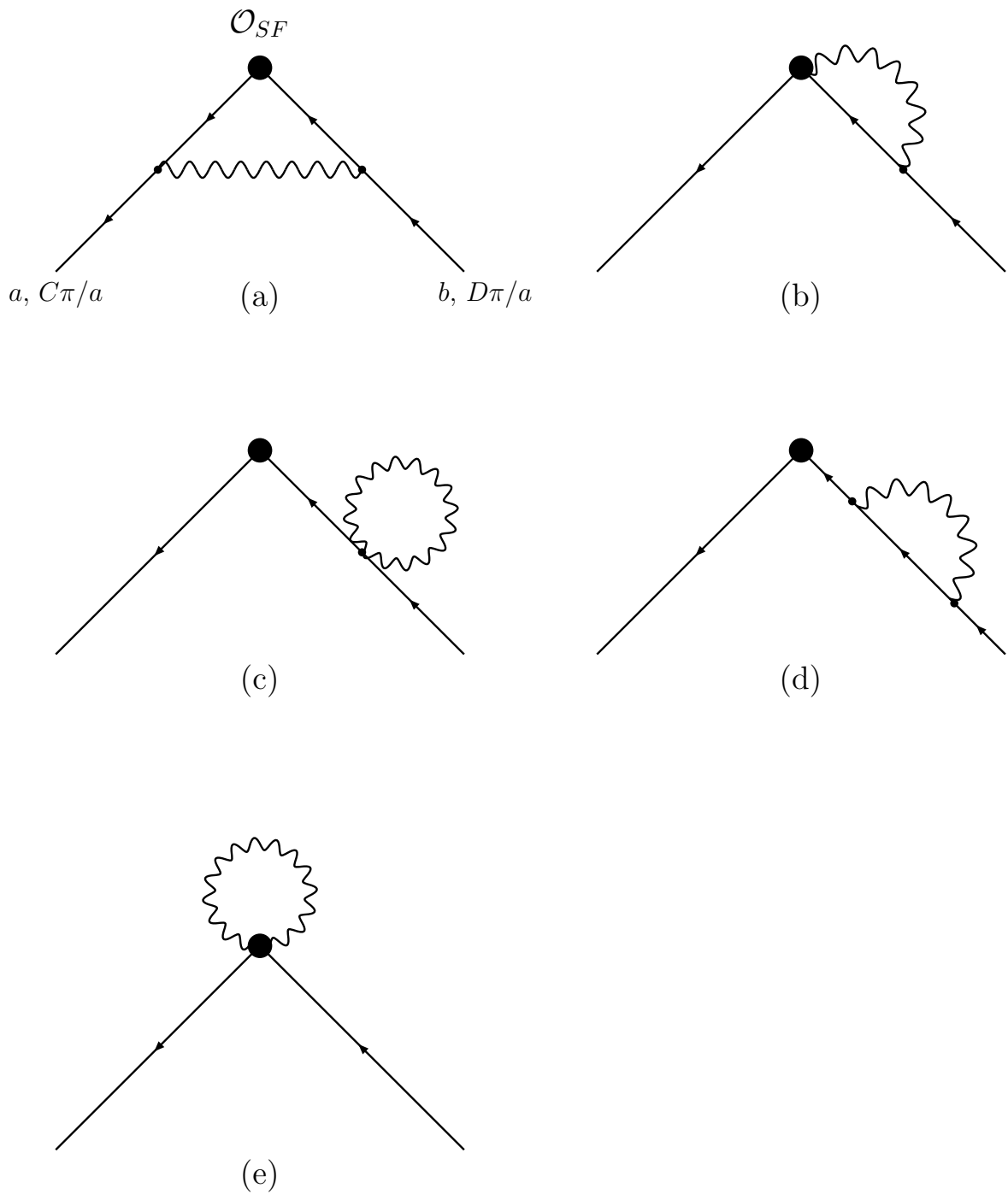
One-loop diagrams for bilinear operators.
Diagrams (b) and (e) are absent for gauge non-invariant operators.
\label{bilineargp}
}
\end{figure}
\newpage
%
%
%
%
%
\newsavebox{\GaT}
\savebox{\GaT}{

}
\newpage
\begin{figure}[p]
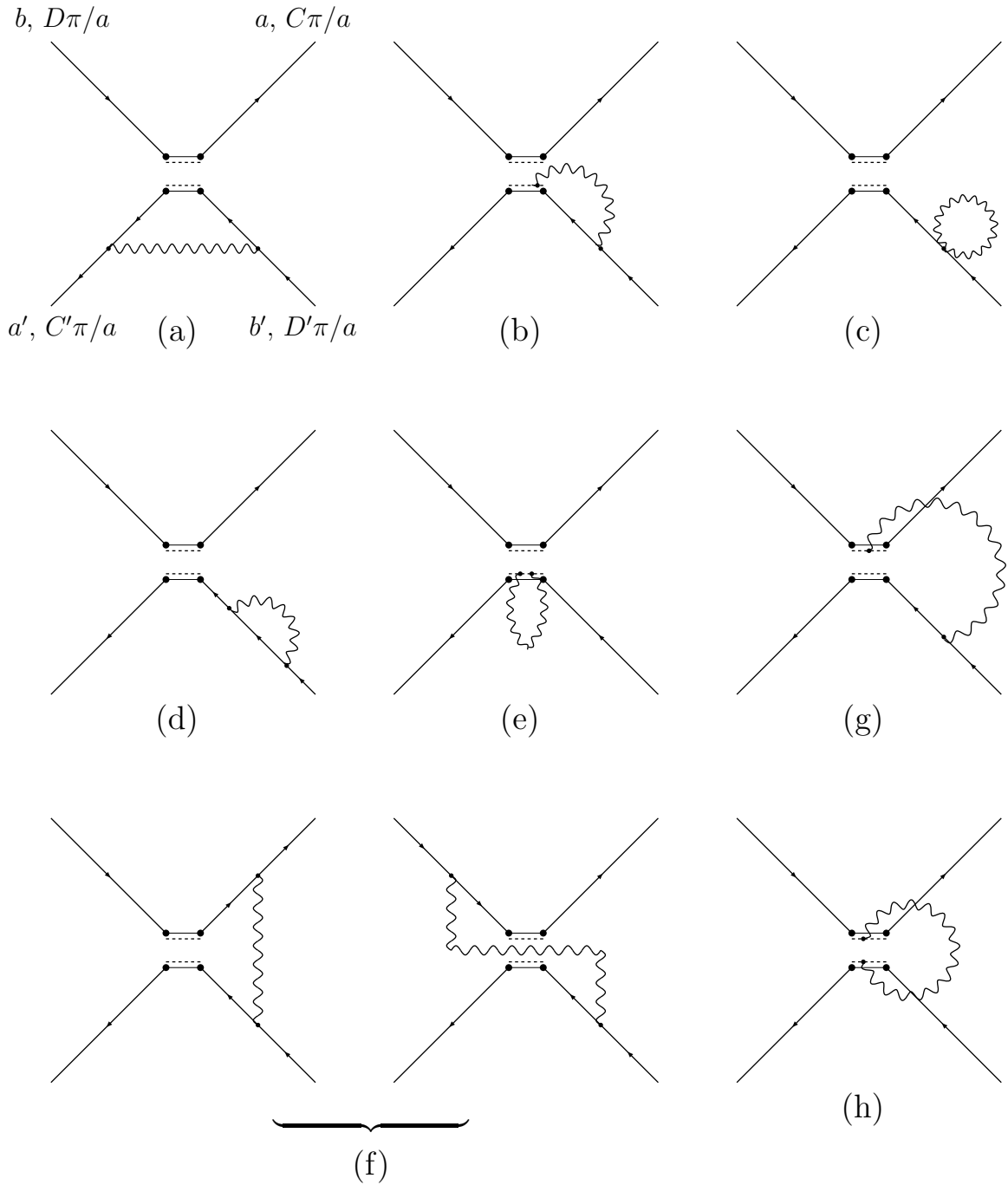

\vspace{-6cm}
\begin{center}
   \begin{tabular}{ccccc}
        \usebox{\GaT}  & $\ \ $ & \usebox{\GbT}  & $\ \ $ & \usebox{\GcT}     \\
        {\large (a)}   & $\ \ $ & {\large (b)}   & $\ \ $ & {\large (c)}      \\
        \usebox{\GdT}  & $\ \ $ & \usebox{\GeT}  & $\ \ $ & \usebox{\GgT}     \\
        {\large (d)}   & $\ \ $ & {\large (e)}   & $\ \ $ & {\large (g)}      \\
        \usebox{\GfaT} & $\ \ $ & \usebox{\GfbT} & $\ \ $ & \usebox{\GhT}     \\
        \multicolumn{4}{c}{$\underbrace{\hspace{3cm}}_{} $} & {\large (h)} \\
        \multicolumn{4}{c}{ {\large (f)} }\\
   \end{tabular}
\end{center}
\caption{
One-loop diagrams for color two-loop four-quark operators
$\bar{Q}^a (\gamma_{S } \otimes\xi_{F })Q^a \cdot \bar{Q}^b (\gamma_{S'} \otimes\xi_{F'})Q^b$.
Thick horizontal bars at the four-quark vertices signify contraction of spin-flavor
quantum numbers,
while dotted lines represent link factors and flow of color indices.
\label{fourfermigp}
}
\end{figure}
%
%
%
%
%
\newsavebox{\GaO}
\savebox{\GaO}{

}
\newpage
\begin{figure}[p]
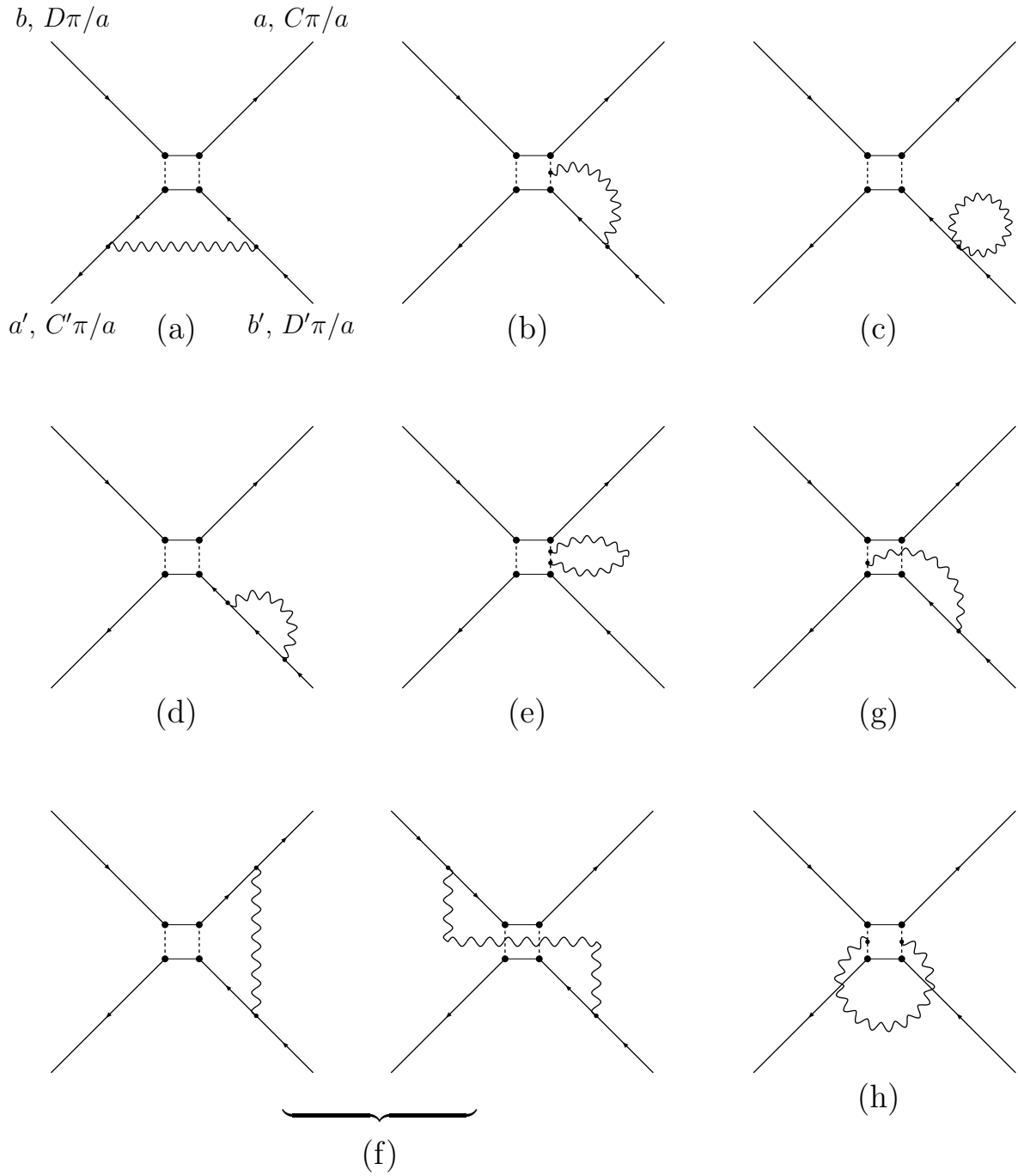

\vspace{-6cm}
\begin{center}
   \begin{tabular}{ccccc}
        \usebox{\GaO}  & $\ \ $ & \usebox{\GbO}  & $\ \ $ & \usebox{\GcO}     \\
        {\large (a)}   & $\ \ $ & {\large (b)}   & $\ \ $ & {\large (c)}      \\
        \usebox{\GdO}  & $\ \ $ & \usebox{\GeO}  & $\ \ $ & \usebox{\GgO}     \\
        {\large (d)}   & $\ \ $ & {\large (e)}   & $\ \ $ & {\large (g)}      \\
        \usebox{\GfaO} & $\ \ $ & \usebox{\GfbO} & $\ \ $ & \usebox{\GhO}     \\
        \multicolumn{4}{c}{$\underbrace{\hspace{3cm}}_{} $} & {\large (h)} \\
        \multicolumn{4}{c}{ {\large (f)} }\\
   \end{tabular}
\end{center}
\caption{
One-loop diagrams for color one-loop four-quark operators
$\bar{Q}^a (\gamma_{S } \otimes\xi_{F })Q^b \cdot \bar{Q}^b (\gamma_{S'} \otimes\xi_{F'})Q^a$.
The meaning of the symbols is the same in Fig.~{\protect \ref{fourfermigp}}.
\label{fourfermigpone}
}
\end{figure}
\end{document}